\newif\ifdraft
\newif\iffull
\newif\ifcomment
\newif\iflatexdiff
\newif\ifbibtex
\newif\ifpreprint
\latexdifftrue   %for latex diff
\bibtextrue      %generation of references (off for submission)
\preprinttrue    %version for arxiv
\fulltrue        %includes author and acknoledgement
\def\dvers{v0.9}
\def\snntitle{$\snn$}
\ifpreprint
\def\snntitle{$\snnbf$}
\fi
\def\dtitle{Forward-central two-particle correlations\\in p--Pb collisions at \snntitle\ = 5.02 TeV}
\def\stitle{Forward-central two-particle correlations in p--Pb collisions}
\ifpreprint
\documentclass[ALICE,manyauthors,12pt]{cernphprep}
\usepackage[comma,square,numbers,sort&compress]{natbib}
\usepackage{hyperref}
\else
\documentclass[final,3p,12pt]{elsarticle}
\biboptions{comma,square,numbers,sort&compress}
\usepackage{hyperref}
\fi
\usepackage{graphicx}  % needed for figures
\usepackage{dcolumn}   % needed for some tables
\usepackage{multirow}
\usepackage{bm}        % for math
\usepackage{amssymb}   % for math
\usepackage{amsfonts}
\usepackage{graphics}
\usepackage{grffile}   % to handle dots in graphics file names
\usepackage{epsfig}
\usepackage{units}
\usepackage[usenames]{color}
\usepackage[normalem]{ulem} % for strikethroughs (\sout{})
\usepackage{color}
\usepackage[utf8]{inputenc}
\usepackage[T1]{fontenc}
\iflatexdiff
\RequirePackage{color}\definecolor{RED}{rgb}{1,0,0}\definecolor{BLUE}{rgb}{0,0,1}

\fi
\ifpreprint
\else
 
\fi
\newcommand{\gevc}         {\ensuremath{\mathrm{GeV}/c}}
\newcommand{\mevc}         {\ensuremath{\mathrm{MeV}/c}}

\newcommand{\vtwo}         {\ensuremath{v_{2}\{ \mathrm{2PC, sub} \} }}
\newcommand{\vnsub}        {\ensuremath{v_{n}\{ \mathrm{2PC, sub} \} }}

\newcommand{\vtwomu}       {\ensuremath{v_{2}^{\mu} \{ \mathrm{2PC, sub} \} }}

\newcommand{\ptt}          {\ensuremath{p_{\mathrm{T}}^{\mathrm t}}}
\newcommand{\pta}          {\ensuremath{p_{\mathrm{T}}^{\mathrm a}}}

\newcommand{\ZNA}          {\rm{ZNA}}
\newcommand{\ZNC}          {\rm{ZNC}}

\newcommand{\TPC}          {\rm{TPC}}
\newcommand{\VZERO}        {\rm{V0}}
\newcommand{\VZEROA}       {\rm{V0-A}}
\newcommand{\VZEROC}       {\rm{V0-C}}
\newcommand{\muonarm}      {\rm{muon spectrometer}}

\newcommand{\pp}           {pp}

\newcommand{\Aa}           {\mbox{A--A}}
\newcommand{\PbPb}         {\mbox{Pb--Pb}}

\newcommand{\pPb}          {\mbox{p--Pb}}

\newcommand{\Pbp}          {\mbox{Pb--p}}
\newcommand{\dAu}          {\mbox{d--Au}}

\newcommand{\dNdeta}       {\mathrm{d}N_\mathrm{ch}/\mathrm{d}\eta}

\newcommand{\s}            {\ensuremath{\sqrt{s}}}
\newcommand{\pt}           {\ensuremath{p_{\mathrm{T}}}}

\newcommand{\snn}          {\ensuremath{\sqrt{s_{\mathrm{NN}}}}}
\newcommand{\snnbf}        {\ensuremath{\mathbf{{\sqrt{\textit{s}_{\mathbf NN}}}}}}

\newcommand{\avg}[1]       {\ensuremath{\left\langle#1\right\rangle}}

\newcommand{\dd}           {\ensuremath{\mathrm{d}}}

\newcommand{\Dphi}         {\ensuremath{\Delta\varphi}}
\newcommand{\Deta}         {\ensuremath{\Delta\eta}}
\newcommand{\Ntrig}        {\ensuremath{N_{\mathrm{trig}}}}

\newcommand{\dNassoc}      {\ensuremath{\frac{\dd^2N_{\mathrm{assoc}}}{\dd\Deta\dd\Dphi}}}

\newcommand{\Tab}[1]       {Tab.~\ref{#1}}
\newcommand{\Fig}[1]       {Fig.~\ref{#1}}

\newcommand{\Sect}[1]      {Sec.~\ref{#1}}

\newcommand{\Eq}[1]        {Eq.~\ref{#1}}

\newcommand{\Ref}[1]       {Ref.~\cite{#1}}
\newcommand{\Refs}[1]      {Refs.~\cite{#1}}

\newcommand{\red}[1]       {\textcolor{red}{#1}}

\newcommand{\warn}[1]      {{\small\textbf{\red{(!}\footnote{\textbf{\red{(!)}}~#1}\red{)}}}\marginpar{\textbf{\red{---}}}}

\newcommand{\com}[1]       {}

\newcommand{\zvtx}         {\ensuremath{z_{\mathrm{vtx}}}}
\newcommand{\MB}           {\ensuremath{\mathrm{MB}}}

\newcommand{\mulpt}       {\ensuremath{\mu}-low-\ensuremath{\pt}}
\newcommand{\muhpt}       {\ensuremath{\mu}-high-\ensuremath{\pt}}
\iflatexdiff

\renewcommand{\xout}[1]    {\textcolor{red}{\sout{#1}}}
 
\else

\renewcommand{\xout}[1]    {}
\fi
\newcommand{\hide}[1]      {\textcolor{white}{#1}}

\graphicspath{{./img/}}
\ifdraft
\usepackage{lineno}
\linenumbers
\modulolinenumbers[5]
\usepackage{fancyhdr}
\else
\renewcommand{\warn}[1]{}

\fi
\begin{document}
%==========================================================%
\newlength{\figlen}
\setlength{\figlen}{\linewidth}
\ifpreprint
\setlength{\figlen}{0.95\linewidth}
\begin{titlepage}
\PHyear{2015}
\PHnumber{155}                   % required, obtained from PH
\PHdate{22 June}             % required

\title{\dtitle}
\ShortTitle{\stitle}
\Collaboration{ALICE Collaboration%
         \thanks{See Appendix~\ref{app:collab} for the list of collaboration members}}
\ShortAuthor{ALICE Collaboration} % appears on left page headers, do not change
\ifdraft
\begin{center}
\today\\ \color{red}DRAFT \dvers\ \hspace{0.3cm} \$Revision: 1082 $\color{white}:$\$\color{black}\vspace{0.3cm}
\end{center}
\fi
\else
\begin{frontmatter}
\title{\dtitle}
\iffull
\input{authors-plb.tex}
\else
\ifdraft
\author{ALICE Collaboration \\ \vspace{0.3cm} 
\today\\ \color{red}DRAFT \dvers\ \hspace{0.3cm} \$Revision: 1082 $\color{white}:$\$\color{black}}
\else
\author{ALICE Collaboration}
\fi
\fi
\fi
%==========================================================%
\begin{abstract}
Two-particle angular correlations between trigger particles in the forward pseudorapidity range ($2.5 < |\eta| < 4.0$) 
and associated particles in the central range~($|\eta| < 1.0$) are measured with the ALICE detector in \pPb\ collisions 
at a nucleon--nucleon centre-of-mass energy of \unit[5.02]{TeV}.
The trigger particles are reconstructed using the muon spectrometer, and the associated particles 
by the central barrel tracking detectors.
In high-multiplicity events, the double-ridge structure, previously discovered in two-particle angular correlations at midrapidity, 
is found to persist to the pseudorapidity ranges studied in this Letter.
The second-order Fourier coefficients for muons in high-multiplicity events are extracted after jet-like correlations 
from low-multiplicity events have been subtracted. 
The coefficients are found to have a similar transverse momentum~($\pt$) dependence in p-going (\pPb) and Pb-going (\Pbp) configurations, 
with the Pb-going coefficients larger by about $16\pm6$\%, rather independent of $\pt$ within the uncertainties of the measurement. 
The data are compared with calculations using the AMPT model, which predicts a different
$\pt$ and $\eta$ dependence than observed in the data.
The results are sensitive to the parent particle $v_2$ and composition of reconstructed muon tracks, 
where the contribution from heavy flavour decays is expected to dominate at $\pt>2$ GeV/$c$.

\ifdraft 
\ifpreprint
\end{abstract}
\end{titlepage}
\else
\end{abstract}
%\begin{keyword}
%\end{keyword}
\end{frontmatter}
%\maketitle
\newpage
\fi
\fi
\ifdraft
\thispagestyle{fancyplain}
\else
\end{abstract}
\ifpreprint
\end{titlepage}
\else
\end{frontmatter}
\fi
\fi
\setcounter{page}{2}

%==========================================================%
%==========================MAIN============================%
%==========================================================%

% $Id: fmc_intro.tex 1073 2015-06-25 11:16:25Z loizides $

%%%%%%%%%%%%%%%%%%%%%%%%%%%%%%%%%%%%%%%%%%%%%%%%%%%%%%%%%%%%%%%%%%%%%%%%%%%%%%%%%%%%%%%%%%%%%%%%%%%%
\section{Introduction}
\label{sec:intro}
%%%%%%%%%%%%%%%%%%%%%%%%%%%%%%%%%%%%%%%%%%%%%%%%%%%%%%%%%%%%%%%%%%%%%%%%%%%%%%%%%%%%%%%%%%%%%%%%%%%%
Measurements of correlations in $\Dphi$ and $\Deta$,
where $\Dphi$ and $\Deta$ are the differences in azimuthal angle~($\varphi$) and pseudorapidity~($\eta$) 
between two particles, respectively, provide insight on the underlying mechanism of particle production 
in collisions of hadrons and nuclei at high energy.

For such measurements in proton--proton~(\pp) collisions, jet production leads to a characteristic 
peak-like structure on the ``near side'' (at $\Dphi \approx 0$, $\Deta \approx 0$) and an elongated structure in $\Deta$ 
on the ``away side'' (at $\Dphi \approx \pi$)~\cite{Wang:1992db}.
In nucleus–nucleus~(\Aa) collisions, ridge-like structures extending over a long range along the $\Deta$ axis 
emerge on the near and away sides, in addition to the jet-related correlations~\cite{Adams:2004pa, Alver:2008gk, Alver:2009id, Abelev:2009qa, 
Chatrchyan:2011eka, Aamodt:2011by, Agakishiev:2011pe, Chatrchyan:2012wg, ATLAS:2012at, Aamodt:2011vk, Adare:2011tg, Abelev:2012di, Chatrchyan:2012vqa}.
The Fourier decomposition of the correlation in $\Dphi$ at large $\Deta$ is dominated by the second- and third-order 
harmonic coefficients $v_2$ and $v_3$, but significant harmonics have been measured up to $v_6$~\cite{Aamodt:2011vk, Adare:2011tg, 
Aamodt:2011by, Chatrchyan:2011eka, Chatrchyan:2012wg, Abelev:2012di, Chatrchyan:2012vqa, ATLAS:2012at,Chatrchyan:2013kba,Aad:2014vba}.  
In \Aa\ collisions, the $v_n$ coefficients are interpreted as the collective response of the created matter to the collision geometry 
and fluctuations in the initial state~\cite{Ollitrault:1992bk,Alver:2010gr}, and are used to extract its transport properties
in hydrodynamic models~\cite{Alver:2010dn, Schenke:2010rr, Qiu:2011hf}.

Long-range \com{($2<|\Deta|<4$)}ridge structures on the near side ($\Dphi \approx 0$) were also observed in 
high-multiplicity \pp\ collisions at a centre-of-mass energy $\s=$~\unit[7]{TeV}~\cite{Khachatryan:2010gv} and in proton--lead~(\pPb) 
collisions at a nucleon--nucleon centre-of-mass energy $\snn=$ \unit[5.02]{TeV}~\cite{CMS:2012qk}. 
Shortly after, measurements in which the contributions from jet fragmentation were suppressed by subtracting 
the correlations extracted from low-multiplicity events, revealed the presence of essentially the same long-range structures on 
the away side as on the near side in high-multiplicity events~\cite{Abelev:2012ola,Aad:2012gla}.  
Evidence of long-range double-ridge structures in high-multiplicity deuteron--gold~(\dAu) collisions at 
$\snn=$~\unit[0.2]{TeV} was also reported~\cite{Adare:2013piz}.  
By now, the existence of long-range correlations in \pPb\ collisions is firmly established by measurements~\cite{Aad:2013fja,Chatrchyan:2013nka,
Abelev:2014mda,Aad:2014lta,Khachatryan:2015waa} involving four, six or more particle correlations, with the lower-order correlations 
removed~\cite{Bilandzic:2010jr}, demonstrating that the long-range ridges originate from genuine multi-particle correlations.
Intriguingly, the transverse momentum  dependence of the extracted $v_n$~\cite{Aad:2013fja,Chatrchyan:2013nka,Aad:2014lta}, 
and the particle-mass dependence of $v_n$~\cite{Abelev:2013wsa,Adare:2014keg,Khachatryan:2014jra} are found to be qualitatively similar 
to those measured in \Aa\ collisions.

The similarity of the ridges in the \pp, \pPb, \dAu\  and \Aa\ systems suggests the possibility of a common hydrodynamical origin~\cite{Avsar:2010rf,
Werner:2010ss,Deng:2011at,Bozek:2011if,Bozek:2012gr,Bozek:2013uha,Shuryak:2013ke,Basar:2013hea}. 
However, whether hydrodynamical models can indeed be reliably applied to such small systems is under intense debate~\cite{Bzdak:2013zma}.
Other proposed mechanisms involve initial-state effects, such as gluon saturation and extended color connections forming along the longitudinal
direction~\cite{Dusling:2012cg,Dusling:2012wy,Kovchegov:2012nd,Dusling:2013oia,Dumitru:2013tja} 
or final-state parton--parton induced interactions~\cite{Arbuzov:2011yr,Wong:2011qr,Strikman:2011cx,Alderweireldt:2012kt,Bzdak:2014dia}.

Further insight into the production mechanism of these long-range correlation structures may be gained by studying their $\eta$-dependence.
A preliminary result~\cite{CMS-PAS-HIN-14-008}\com{ from the CMS collaboration} indicates a mild $\eta$ dependence, 
but the measurement is limited to \mbox{$|\eta|<2$}.
A similar magnitude of the two-particle correlation amplitudes in the Au-going and d-going directions at $2.8<|\eta|<3.8$ has also been reported 
in \dAu\ collisions at $\snn=$~\unit[0.2]{TeV}\com{ by the STAR collaboration}~\cite{Adamczyk:2015xjc}. %
Calculations for $v_2$ at large $\eta$~($2.5<|\eta|<4$) in \pPb\ collisions at $\snn=$~\unit[5.02]{TeV} from a 3+1~dimensional, viscous hydrodynamical 
model\com{\cite{Bozek:2011if}} and a multi-phase transport model~(AMPT)\com{\cite{Bzdak:2014dia}} predict a stronger $\eta$ dependence, 
with about 50\% and 30\% larger $v_2$ values on the lead nucleus side for the hydrodynamical and AMPT model, respectively~\cite{Bozek:2015swa}.

In this Letter, we report a measurement of angular correlations between trigger particles in the pseudorapidity range $2.5 < |\eta| < 4.0$
and associated particles in the central range $|\eta| < 1.0$ in \pPb\ collisions at $\snn=$\unit[5.02]{TeV} at the Large Hadron Collider~(LHC). 
The trigger particles are inclusive muons, reconstructed using the ALICE muon spectrometer, and the associated particles are charged particles,
reconstructed by the ALICE central barrel tracking detectors.
As in previous measurements~\cite{Abelev:2012ola,Abelev:2013wsa}, the double ridge is extracted by subtracting the correlations obtained 
in low-multiplicity events from those in high-multiplicity events.
Results for the second order Fourier coefficient for muons, $\vtwomu$, and the ratio of $\vtwomu$ coefficients\footnote{Here, and in the following,
``2PC'' stands for ``two-particle correlation'' and ``sub'' for ``subtraction'', and indicates the analysis technique with which the coefficients are measured.} 
in the Pb-going~(\Pbp) and p-going~(\pPb) directions are reported for high-multiplicity events, and compared to model predictions.
The remainder of the Letter is structured as follows:
We describe the experimental setup in \Sect{sec:setup}, the event and track selection in \Sect{sec:selection},
the analysis method in \Sect{sec:analysis} and the evaluation of the systematic uncertainties in \Sect{sec:systematics}.
Finally, in \Sect{sec:results} we report the results, and compare them with model predictions.
In \Sect{sec:summary} we conclude with a summary. %of the Letter. 

% $Id: fmc_setup.tex 1079 2015-06-25 18:10:06Z ekryshen $

%%%%%%%%%%%%%%%%%%%%%%%%%%%%%%%%%%%%%%%%%%%%%%%%%%%%%%%%%%%%%%%%%%%%%%%%%%%%%%%%%%%%%%%%%%%%%%%%%%%%
\section{Experimental setup}
\label{sec:setup}
%%%%%%%%%%%%%%%%%%%%%%%%%%%%%%%%%%%%%%%%%%%%%%%%%%%%%%%%%%%%%%%%%%%%%%%%%%%%%%%%%%%%%%%%%%%%%%%%%%%%
In 2013, the LHC provided collisions between protons with a beam energy of \unit[4]{TeV} and lead ions with a 
beam energy of \unit[1.58]{TeV} per nucleon, resulting in a centre-of-mass energy of $\snn=\unit[5.02]{TeV}$. 
The beams were set up in two configurations:
a period with the proton momentum in the direction of negative $\eta$ in the ALICE coordinate system, 
denoted as \pPb, followed by a period with reversed beams, denoted as \Pbp.
Due to the asymmetric beam energies, the nucleon-nucleon centre-of-mass reference system moves with a rapidity of 0.465 
in the direction of the proton beam with respect to the ALICE laboratory system.
Pseudorapidity, denoted by $\eta$, is given in the laboratory frame throughout this Letter.

Details on ALICE and its subdetectors can be found in \Refs{Aamodt:2008zz,Abelev:2014ffa}. In the following,
we give a brief summary of the components needed for the measurement reported in the Letter.

Trigger tracks used in this analysis are detected in the \muonarm\ with an acceptance of $-4.0<\eta<-2.5$.
The \muonarm\ consists of a thick absorber of about ten interaction lengths ($\lambda_{\rm I}$), which filters muons 
in front of five tracking stations made of two planes of Cathode Pad Chambers each. 
The third station is placed inside a dipole magnet with a \unit[3]{Tm} integrated field.
The tracking apparatus is completed by a trigger system made of four layers of Resistive Plate Chambers 
placed behind a second absorber of 7.2 $\lambda_{\rm I}$ thickness.
This setup ensures that most of the hadrons in the acceptance are stopped in one of the absorber layers, 
providing a muon purity above 99\% for the tracks used in this analysis.
In \pPb\ collisions, the trigger particle travels in the same direction as the p beam~(p-going case),
while in \Pbp\ collisions in the same direction as the Pb nucleus~(Pb-going case). 

Associated particles in $|\eta|<1.0$ are reconstructed using the combined information from the Inner Tracking System (ITS) 
and the Time Projection Chamber (TPC), which are located inside the ALICE solenoid with a field of \unit[0.5]{T}.
The ITS consists of six layers of silicon detectors: two layers of Silicon Pixel Detector (SPD), 
surrounded by two layers of Silicon Drift Detector (SDD) and two layers of Silicon Strip Detector (SSD).
SPD tracklets, short track segments reconstructed in the two SPD layers alone, are also used as associated particles.

The \VZERO\ detector, consisting of two arrays with 32 scintillator tiles arranged in four rings each,
is used to generate the minimum-bias trigger and offline for multiplicity selection~\cite{Abbas:2013taa}.
The detector covers the full azimuth within $2.8<\eta<5.1$ (\VZEROA) and $-3.7<\eta<-1.7$ (\VZEROC). 
The timing information of the \VZERO\ is also used for offline rejection of interactions of the beam with residual gas.
In addition, two neutron Zero Degree Calorimeters (ZDCs) located at \unit[$+112.5$]{m} (\ZNA) and \unit[$-112.5$]{m} (\ZNC) 
from the interaction point are used in the offline event selection and as an alternative approach to define event-multiplicity classes.

% $Id: fmc_evtrack.tex 1312 2015-11-08 18:40:33Z loizides $

%%%%%%%%%%%%%%%%%%%%%%%%%%%%%%%%%%%%%%%%%%%%%%%%%%%%%%%%%%%%%%%%%%%%%%%%%%%%%%%%%%%%%%%%%%%%%%%%%%%%
\section{Event and track selection}
\label{sec:selection}
%%%%%%%%%%%%%%%%%%%%%%%%%%%%%%%%%%%%%%%%%%%%%%%%%%%%%%%%%%%%%%%%%%%%%%%%%%%%%%%%%%%%%%%%%%%%%%%%%%%%
The online event selection used in this analysis is based on a combination of minimum-bias (\MB) and muon trigger inputs. 
The \MB\ selection uses the coincidence between hits in the \VZEROA\ and \VZEROC\ detectors and covers 99.2\% of the non-single-diffractive 
cross section as described in~\cite{ALICE:2012xs}.
Only approximately 5\% of the MB events contain one or more tracks reconstructed in the \muonarm.
In order to increase the number of recorded events, the presence of at least one muon above a $\pt$ threshold 
was required in addition to the \MB\ trigger condition.
Two different thresholds were used:
a low-\pt\ threshold corresponding to about~\unit[$0.5$]{\gevc}~(\mulpt) 
and a higher $\pt$ threshold corresponding to about~\unit[$4.2$]{\gevc}~(\muhpt). 
These thresholds are not sharp and the reported values correspond to a 50\% trigger probability for a muon candidate. 
The integrated luminosity collected with \muhpt\ triggers is $5.0\ {\rm nb}^{-1}$ in the \pPb\ and $5.8\ {\rm nb}^{-1}$ in the \Pbp\ periods.
The \mulpt\ trigger class was downscaled by a factor $10$--$35$ depending on the data taking conditions, resulting
in an integrated luminosity of $0.28\ {\rm nb}^{-1}$ in the \pPb\ and $0.26\ {\rm nb}^{-1}$ in the \Pbp\ periods.

The \TPC\ and SDD detectors have longer deadtime compared to the \muonarm, the SPD and the \VZERO. 
Therefore, they were read out only in a fraction of \mulpt\ events~(about 25\% in \pPb\ and below 10\% in \Pbp\ collisions).
Both muon-track and muon-tracklet correlation results were measured in the \pPb\ configuration. 
For \Pbp\ collisions, only muon-tracklet correlations could be studied due to the significantly lower number of triggers with the TPC
in the readout.

The primary-vertex position is determined using reconstructed clusters in the SPD detector as described in \Ref{Abelev:2014ffa}. 
Only events with a reconstructed vertex coordinate along the beam direction (\zvtx) within \unit[7]{cm} from the nominal interaction point are selected.
The probability of multiple interactions in the same bunch crossing (pileup) was dependent on the beam conditions and always below 3\%. 
Pileup events are removed by rejecting triggers with more than one reconstructed vertex. 

\begin{table}[t]
\centering
\begin{tabular}{cc}
Event        &$\avg{\dNdeta}|_{|\eta|<0.5}$\\
class        & $\pt>$~\unit[0]{\gevc}\\
\hline
\hide{0}0--20\%  & $35.8 \pm 0.8$\\
20--40\%         & $23.2 \pm 0.5$\\
40--60\%         & $15.8 \pm 0.4$\\
60--100\%        & $ 6.8 \pm 0.2$\\
\end{tabular}
\caption{\label{tab:multclasses}
V0S multiplicity classes as fractions of the analyzed event sample and the corresponding $\avg{\dNdeta}|_{|\eta|<0.5}$. 
The $\avg{\dNdeta}$ values are not corrected for trigger and vertex-reconstruction inefficiencies, 
which are about 4\% for non-single-diffractive events~\cite{ALICE:2012xs}, mainly affecting the
80-100\% lowest mulitiplicity events~\cite{Adam:2014qja}.
Only systematic uncertainties are listed, since the statistical uncertainties are negligible.
}
\end{table}

All events were characterized by their event activity, and sorted into event classes.
As in previous studies~\cite{Abelev:2012ola,Abelev:2013wsa}, the event characterization was based on the signal in the \VZERO\ detectors.
However, unlike before, both beam orientations were investigated in this Letter. 
Therefore, the signals from only two out the four rings of \VZEROA\ and \VZEROC\ detectors were combined to guarantee a more symmetric acceptance.
On the \VZEROA\ side, the two outermost rings with an acceptance of $2.8<\eta<3.9$, while on the \VZEROC\ side the two innermost rings 
with an acceptance of $-3.7<\eta<-2.7$ were used. This combination is called V0S in the following. 
The definition of the event classes as fractions of the analyzed event sample and their corresponding average number of particles at
midrapidity ($\avg{\dNdeta}|_{|\eta|<0.5}$), measured using tracklets as explained below, is given in \Tab{tab:multclasses}. 

Muon tracks are reconstructed in the geometrical acceptance of the muon spectrometer ($-4 < \eta < -2.5$).
The tracks are required to exit the front absorber at a radial distance from the beam axis, $R_{\rm abs}$, in the range $17.6 < R_{\rm abs} < 89.5$ cm 
in order to avoid regions with large material density. 
The muon identification is performed by matching the tracks reconstructed in the tracking chambers with the corresponding track segments in the trigger chambers. 
Beam-gas tracks, which do not point to the interaction vertex, are removed by a selection on the product of the total momentum 
of a given track and its distance to the interaction vertex in the transverse plane. 
In the analysis, muons in the transverse momentum range $0.5<\pt<4$~\gevc\ were considered.

\begin{figure}[t]
\centering
\includegraphics[width=0.49\textwidth]{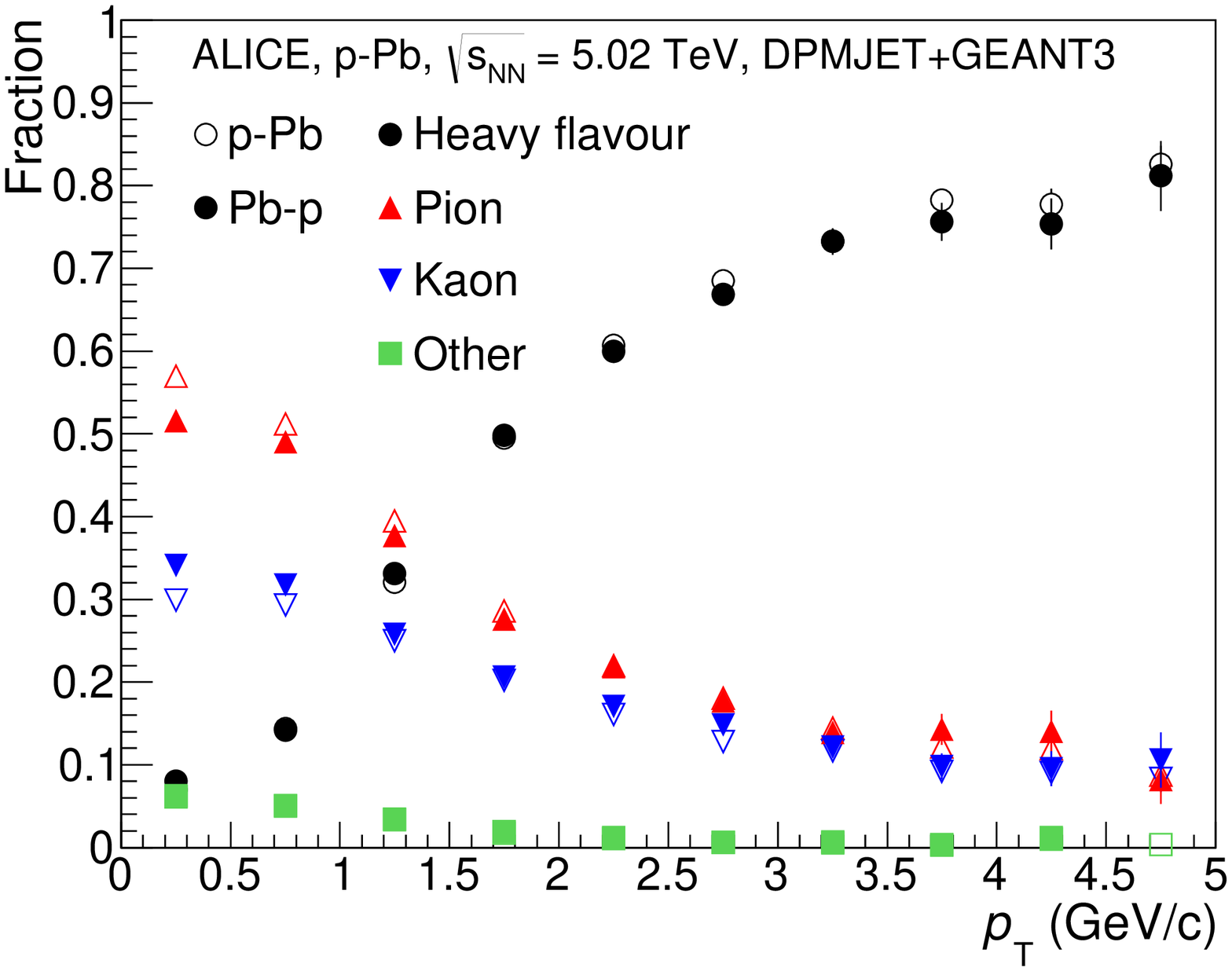}
\hspace{0.cm}
\includegraphics[width=0.49\textwidth]{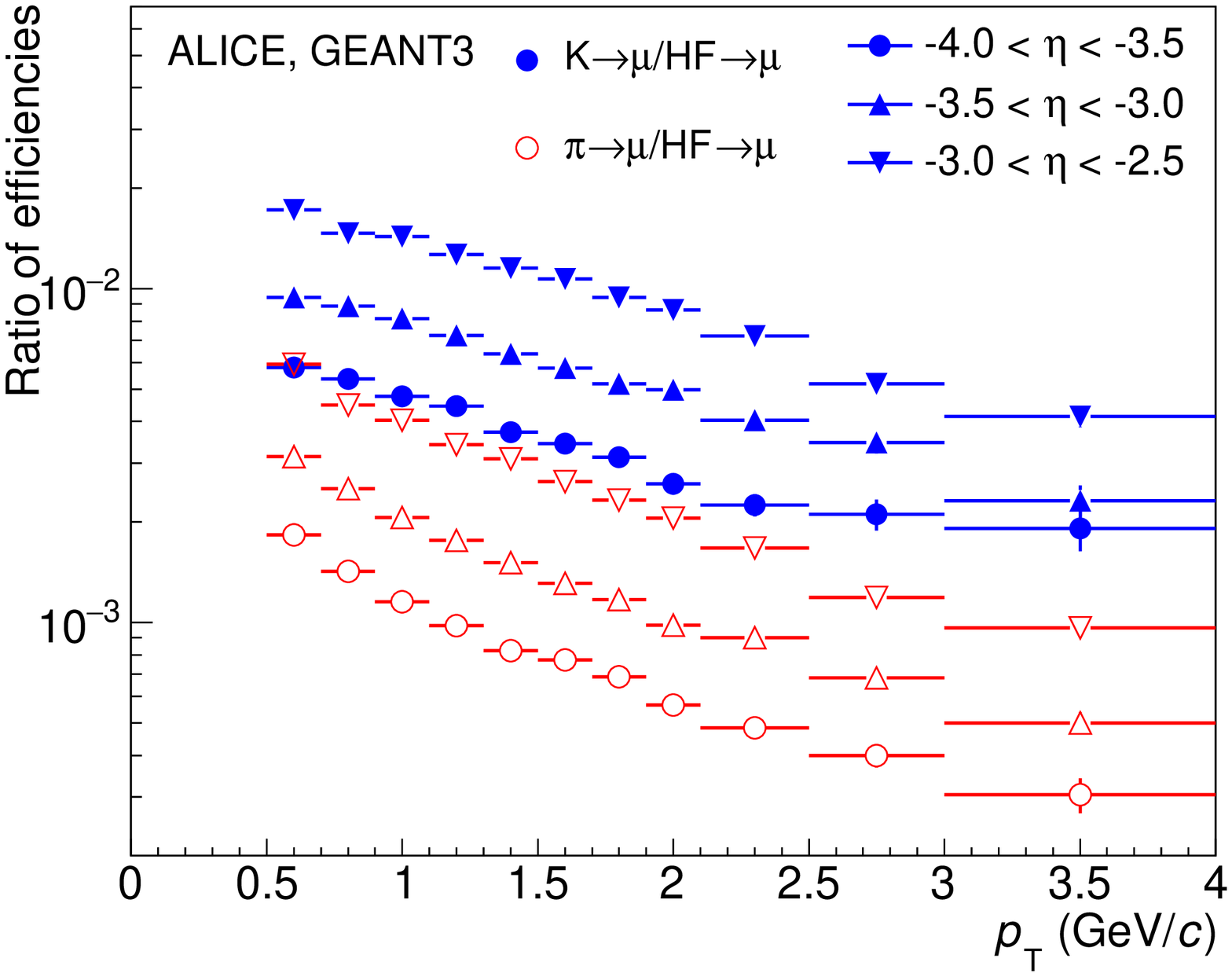}
\caption{\label{fig:mumothers}
 Parent particle composition of reconstructed muon tracks (left panel) and 
 reconstruction efficiency for muons from pion and kaon decays relative to that for heavy flavor~(HF) decay muons (right panel) 
 from a detector simulation of the ALICE muon spectrometer. 
}
\end{figure}

Reconstructed muons mainly originate from weak decays of $\mathrm{\pi}$, $\mathrm{K}$ \footnote{Here, and in the following, pions and kaons refer to the sum of 
both charge states. Neutral particles are also considered in the case of kaons.} and mesons from heavy flavor~(HF) decays.
Because of the different $\pt$ distribution of the various sources and the absorber in front of the spectrometer, which suppresses by design weak decays from light hadrons, 
the parent particle composition for the reconstructed muon tracks changes as a function of $\pt$.
The composition shown as a function of the reconstructed $\pt$ in the left panel of \Fig{fig:mumothers} was evaluated using full detector simulations 
based on the DPMJET Monte Carlo~(MC) event generator~\cite{Roesler:2000he}.
The detector response was simulated using GEANT3 for particle transport~\cite{geant3ref2}.
The composition of parent particles in the simulation differs by less than 10\% for the two beam configurations.
The reconstructed muons are dominated by light-hadron decays below \unit[1.5]{\gevc}, and by heavy flavor decays above \unit[2]{\gevc}.
No significant multiplicity dependence was found.
Similar conclusions are obtained using simulations with the AMPT generator~\cite{Lin:2004en}.

Without strong model assumptions, one cannot deduce the composition of parent particles from the measured muon distribution,
and correct the data for muon decay and absorber effects. 
For comparison of the $v_2$ data with calculations, however, only relative contributions of the parent species matter.
In order to ease future model calculations, the reconstruction efficiencies for muons from pion and kaon decays relative to those for muons from heavy 
flavor decays are provided in the right panel of \Fig{fig:mumothers} as a function of the generated decay muon $\pt$ in different pseudorapidity intervals.
Contributions from muon decays of other particles are significantly smaller than those for pions and can be ignored.
The systematic uncertainty on the relative efficiencies was estimated to be less than $5$\%.

Tracks reconstructed in the ITS and the TPC are selected in the fiducial region $|\eta|<1$ and $0.5 < \pt < 4~\gevc$. 
The track selection used in this Letter is the same as in \Ref{Abelev:2012ola}.

Tracklet candidates are formed using information on the position of the primary vertex and the two hits on the SPD layers~\cite{Aamodt:2010pb}, 
located at a distance of 3.9 and 7.6 cm from the detector centre.
The differences of the azimuthal ($\Delta\varphi_{\,\rm h}$, bending plane) and polar ($\Delta\theta_{\,\rm h}$, non-bending direction) angles of the hits 
with respect to the primary vertex are used to select particles, typically with $\pt>50$~\mevc. 
Particles below 50 \mevc\ are mostly absorbed by material.
Compared to previous analyses~\cite{Aamodt:2010pb,ALICE:2012xs} a tighter cut in $\Delta\varphi_{\,\rm h}$ is applied ($\Delta\varphi_{\,\rm h}$ < \unit[5]{mrad}) 
to select particles with larger $\pt$ and to minimize contributions of fake and secondary tracks to below 2.5\%. 
The corresponding mean $\pt$ of selected particles, estimated from the DPMJET MC, is about \unit[0.75]{\gevc}. 

% $Id: fmc_analysis.tex 1312 2015-11-08 18:40:33Z loizides $

%%%%%%%%%%%%%%%%%%%%%%%%%%%%%%%%%%%%%%%%%%%%%%%%%%%%%%%%%%%%%%%%%%%%%%%%%%%%%%%%%%%%%%%%%%%%%%%%%%%%
\section{Analysis}
\label{sec:analysis}
%%%%%%%%%%%%%%%%%%%%%%%%%%%%%%%%%%%%%%%%%%%%%%%%%%%%%%%%%%%%%%%%%%%%%%%%%%%%%%%%%%%%%%%%%%%%%%%%%%%%
The associated yield of tracks or tracklets per trigger particle in the \muonarm\ 
is measured as a function of the difference in azimuthal angle (\Dphi) and pseudorapidity (\Deta).
As in previous analyses~\cite{Abelev:2012ola,Abelev:2013wsa}, it is defined as
\begin{equation}
Y = \frac{1}{\Ntrig} \dNassoc = \frac{S(\Deta,\Dphi)}{B(\Deta,\Dphi)}, \label{pertriggeryield}
\end{equation}
in intervals of event multiplicity and trigger particle transverse momentum, \ptt.
The variable $\Ntrig$ denotes the total number of trigger particles in the event class and $\ptt$ interval,
not corrected for single-muon efficiency.
The signal distribution $S(\Deta,\Dphi) = 1/\Ntrig \dd^2N_{\rm same} / \dd\Deta\dd\Dphi$
is the associated yield per trigger particle for particle pairs from the same event, obtained in
\unit[1]{cm}-wide intervals of \zvtx.
A correction for pair acceptance and pair efficiency is obtained by dividing by the background distribution 
$B(\Deta,\Dphi) = \alpha\ \dd^2N_{\rm mixed}/\dd\Deta\dd\Dphi$. 
The background distribution is constructed by correlating trigger particles from one event with the associated particles 
from other events within the same event multiplicity class and \unit[1]{cm}-wide $z_{\rm vtx}$ intervals.
The factor $\alpha$ is used to normalize the background distribution to unity in the $\Deta$ region of maximal pair acceptance.
The final per-trigger yield is obtained by calculating the average over the $z_{\rm vtx}$ intervals weighted by \Ntrig.

\begin{figure}[t]
\centering
\includegraphics[width=0.32\textwidth]{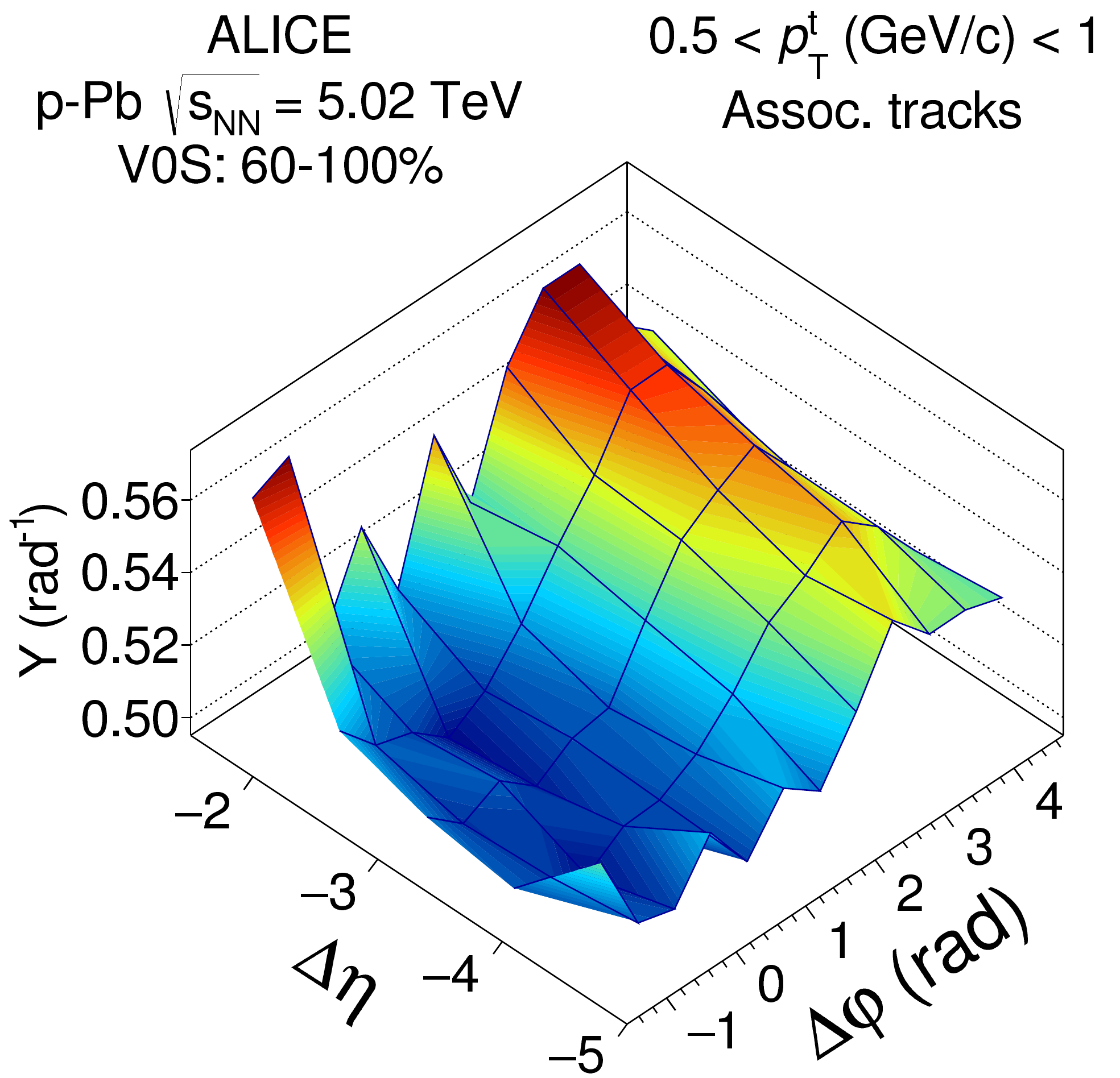}
\includegraphics[width=0.32\textwidth]{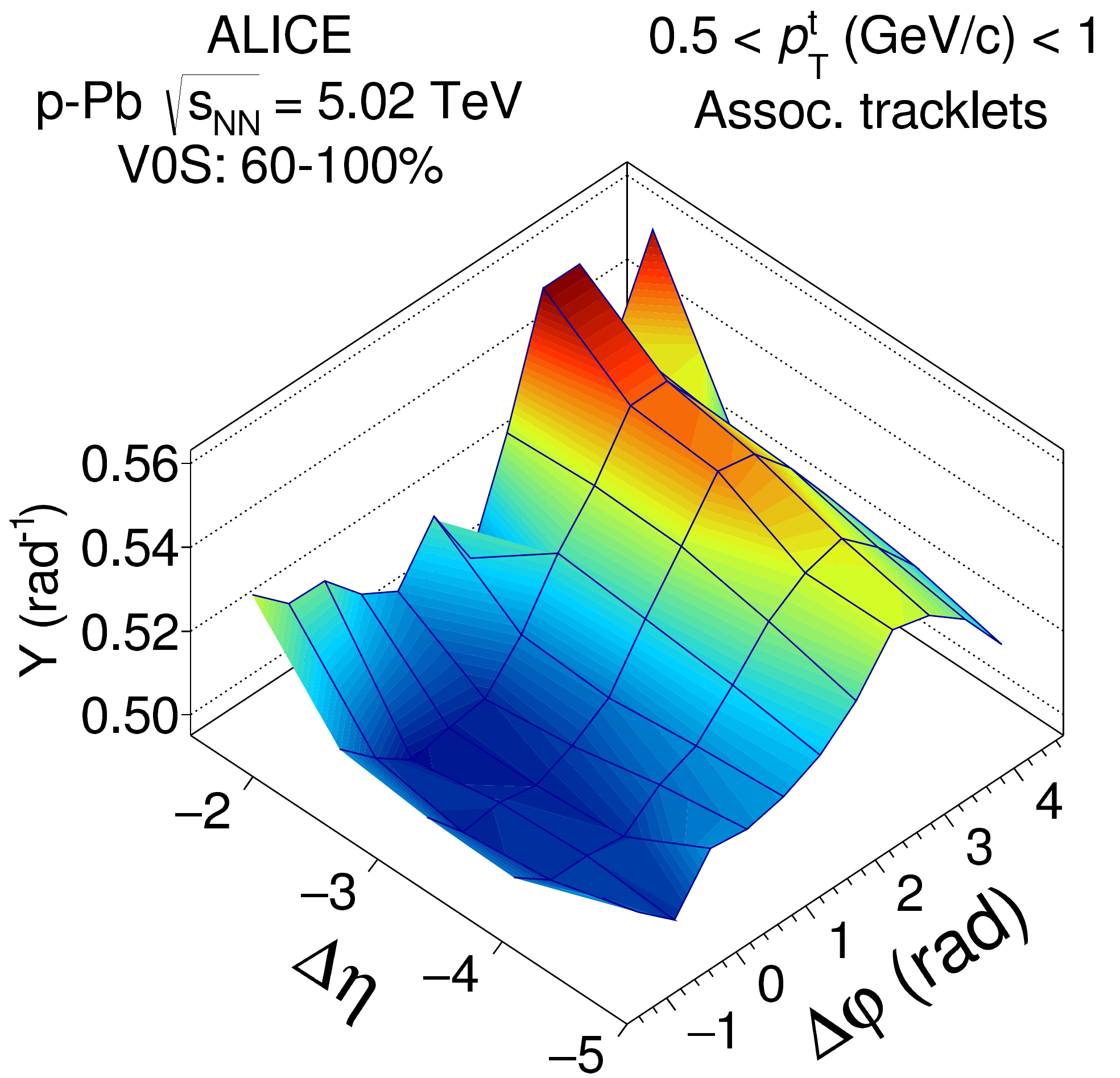}
\includegraphics[width=0.32\textwidth]{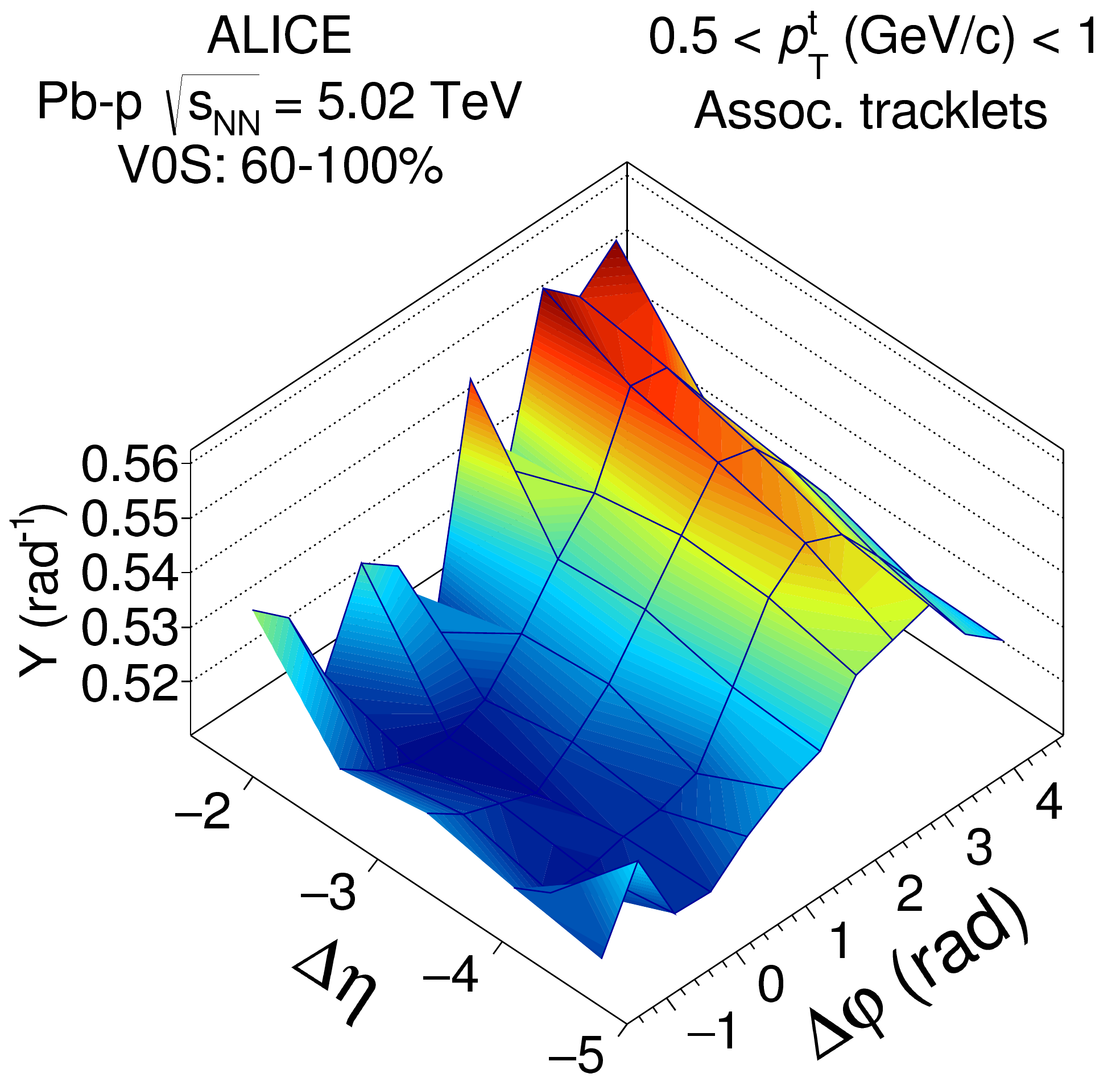}\\
\vspace{0.3cm}
\includegraphics[width=0.32\textwidth]{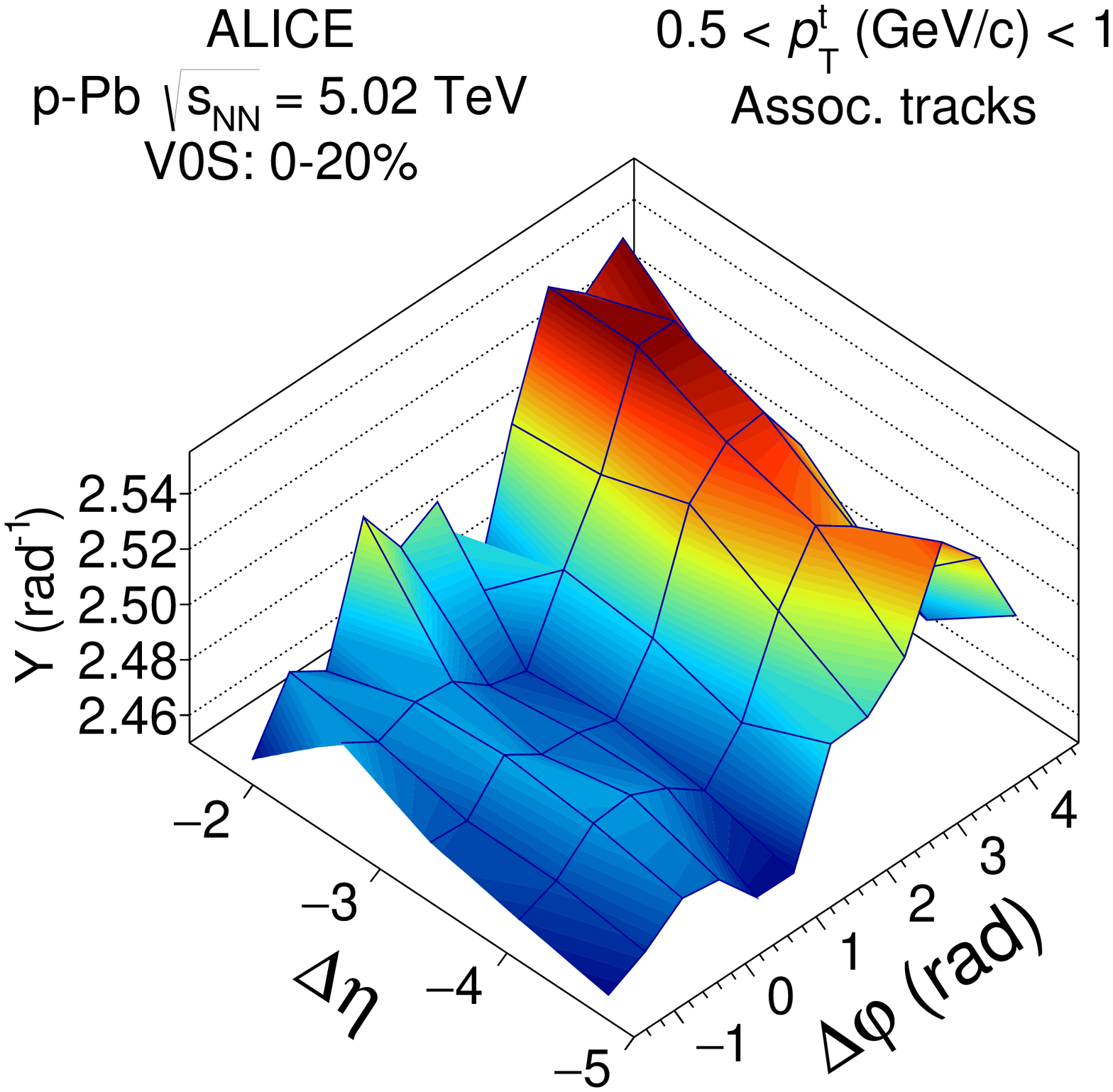}
\includegraphics[width=0.32\textwidth]{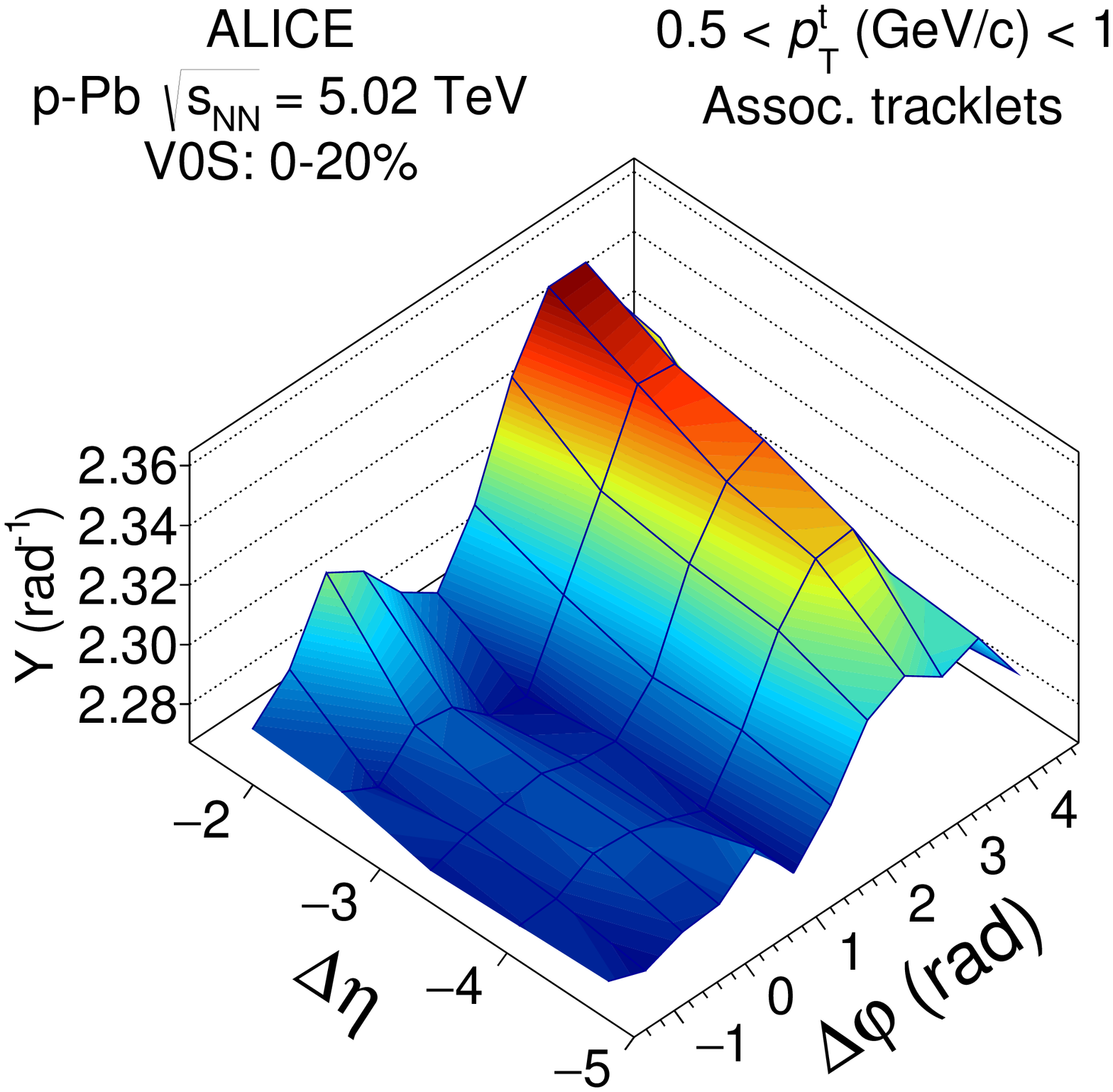}
\includegraphics[width=0.32\textwidth]{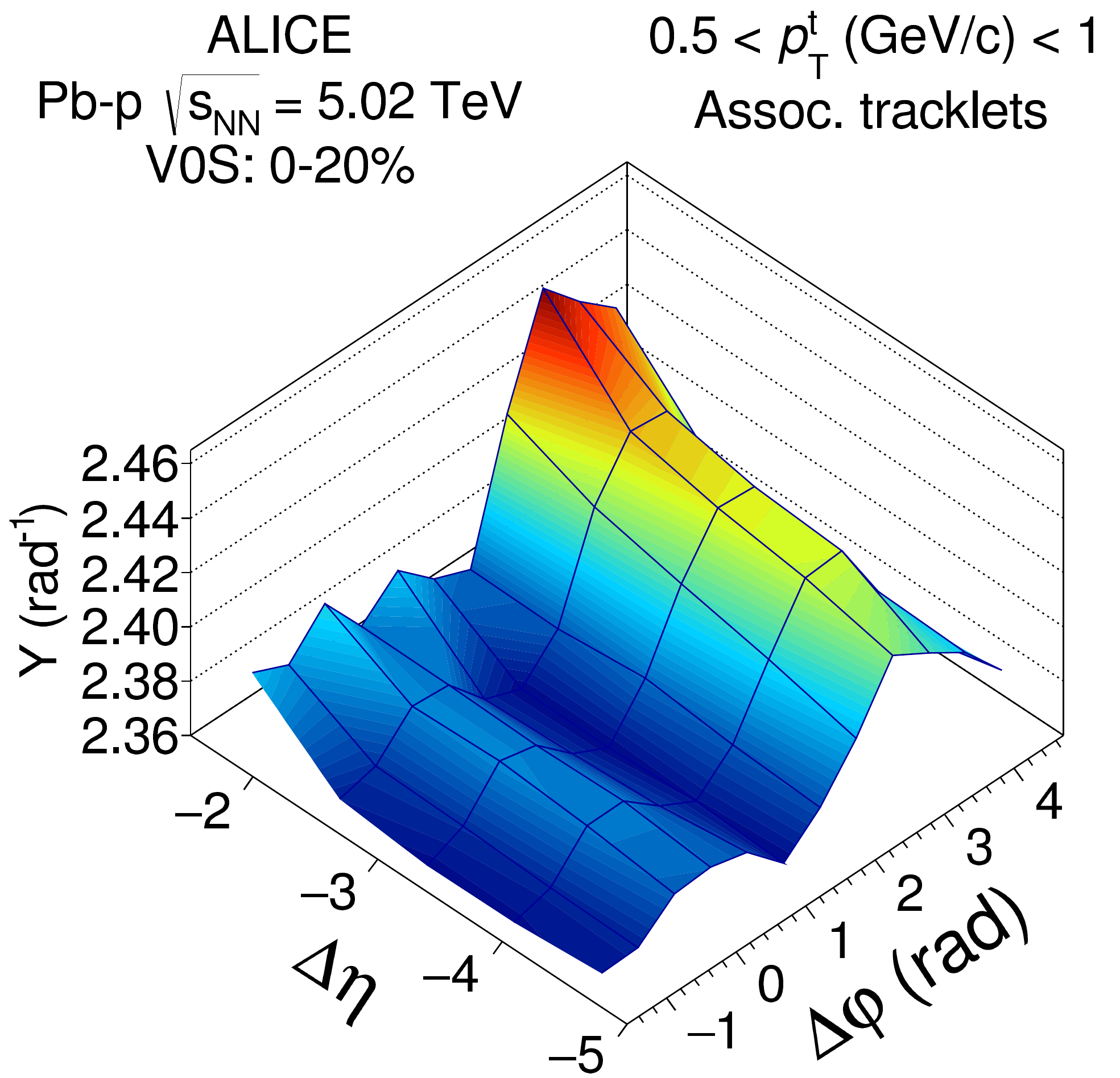}
\caption{\label{fig:corr_before_sub}
 Associated yield per trigger particle as a function of \Deta\ and \Dphi\ for muon-track correlations in \pPb~(left) 
 and muon-tracklet correlations in \pPb~(middle) and \Pbp~(right panels), measured in 60--100\% (top row)  and 0--20\% (bottom row) event classes. 
 The trigger particle (muon) range is $0.5<\ptt<1$~\gevc, the associated particle intervals are $0.5<\pta<4.0$~\gevc\ 
 for tracks and $0<\Delta\varphi_{\,\rm h}<5$~mrad for tracklets.
 Statistical uncertainties are not shown.
}
\end{figure}

In \Fig{fig:corr_before_sub}, the associated yield per trigger particle as a function of \Dphi\ and \Deta\ for muon-track correlations 
in \pPb~(left) and muon-tracklet correlations in \pPb~(middle) and \Pbp~(right panels), measured in 60--100\% (top row) and 0--20\% (bottom row) 
event classes is shown.
In the low-multiplicity class (60--100\%), the dominant feature is the recoil jet on the away side~($\pi/2 < \Dphi < 3\pi/2$).
While in previous two-particle correlation studies at midrapidity~\cite{Abelev:2012ola,Abelev:2013wsa} the away-side jet structure was mostly flat in \Deta, 
from $\Deta=-1.5$ to $\Deta=-5.0$\com{by approximately 3\%} it decreases, as expected considering the kinematics of dijets at large \Deta. 
The near side ($|\Dphi|<\pi/2$) shows almost no structure in \Dphi\ and \Deta, since it is sufficiently separated 
from the near-side jet peak at $(\Dphi,\Deta) = (0,0)$, so that no contribution from jets is expected.
In the high-multiplicity (0--20\%) class, the away-side jet structure is also visible, and the associated yields are considerably higher than for the 
low-multiplicity (60--100\%) class. 
Moreover, in contrast to the low-multiplicity class, a near-side structure emerges, similar to that previously observed at lower pseudorapidities,
revealing that the near-side ridge extends up to pseudorapidity ranges of $2.5<|\eta|<4$. 
%, ie.\ on average to much lower Bjorken-$x$ values than probed earlier.
 
\begin{figure}[t]
\centering
\includegraphics[width=0.32\textwidth]{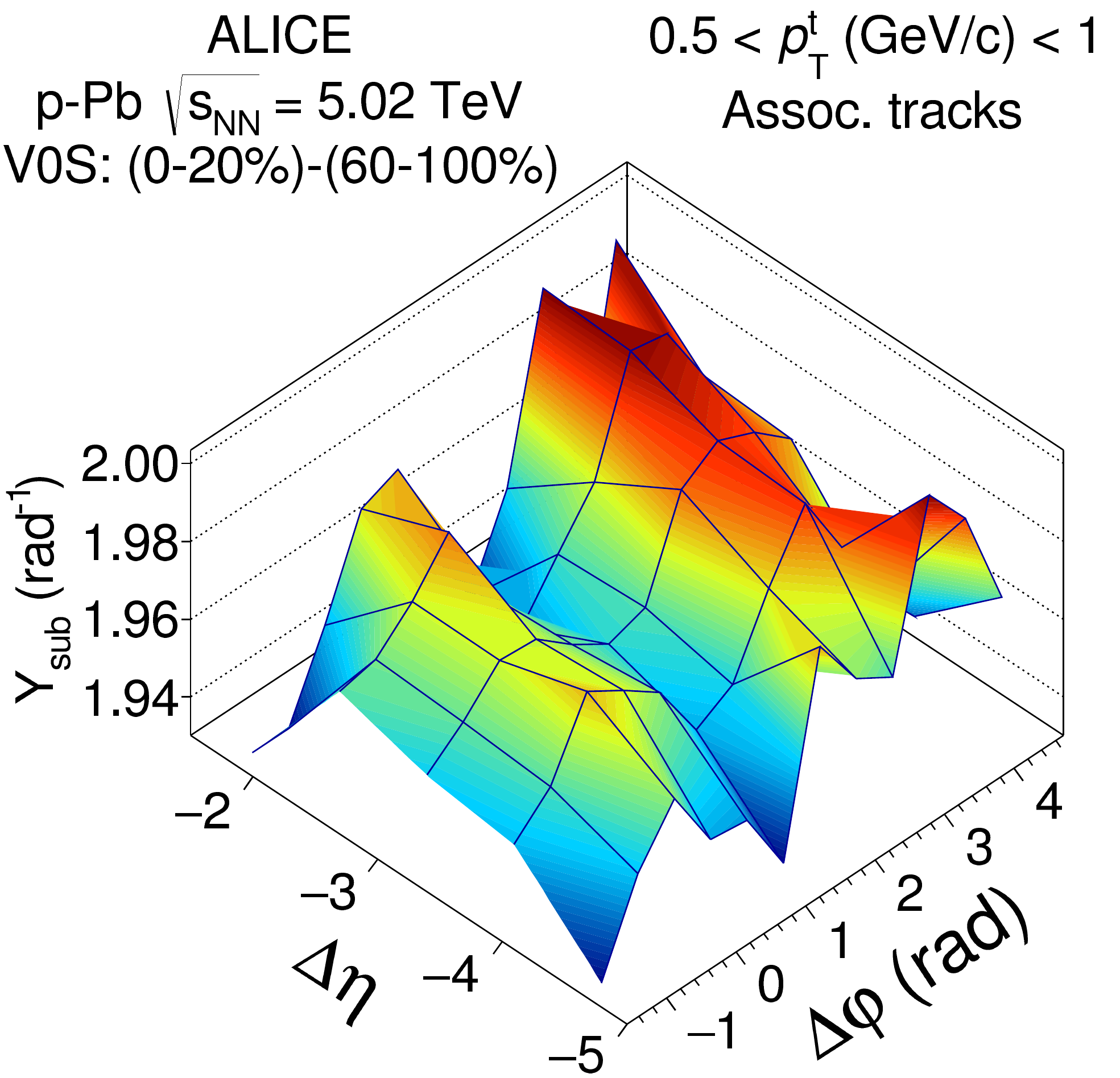}
\includegraphics[width=0.32\textwidth]{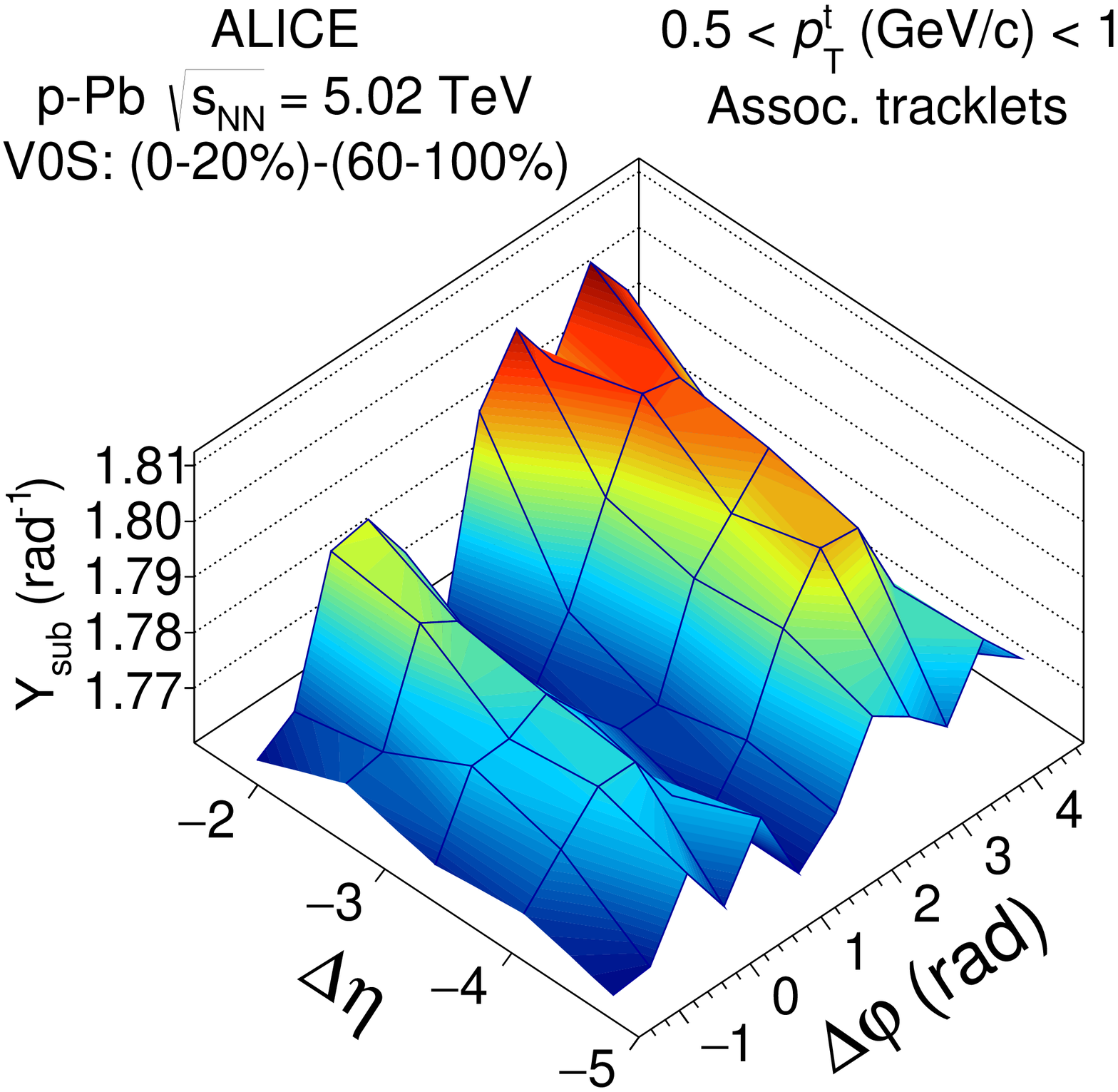}
\includegraphics[width=0.32\textwidth]{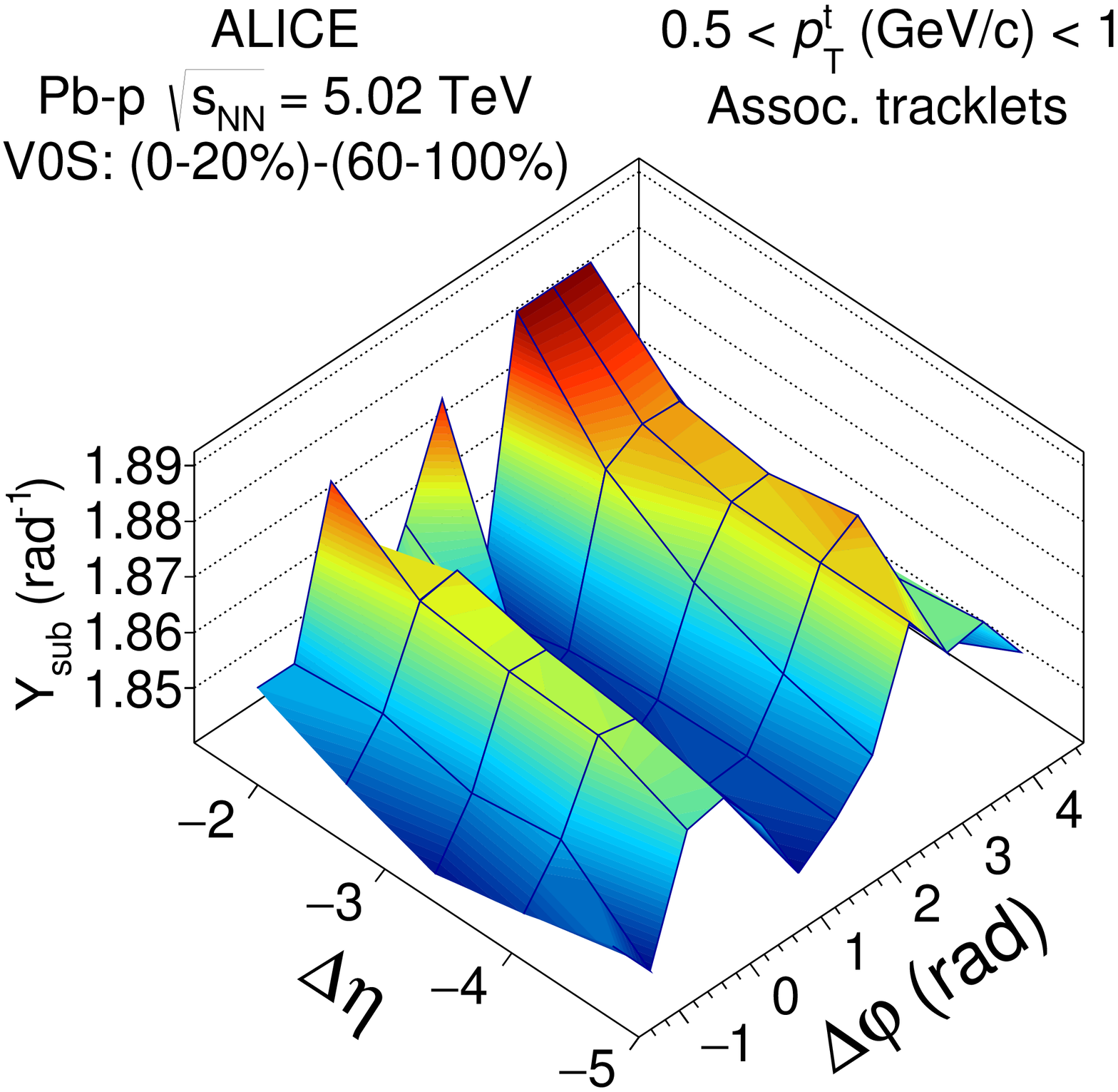}\\
\vspace{0.3cm}
\includegraphics[width=0.32\textwidth]{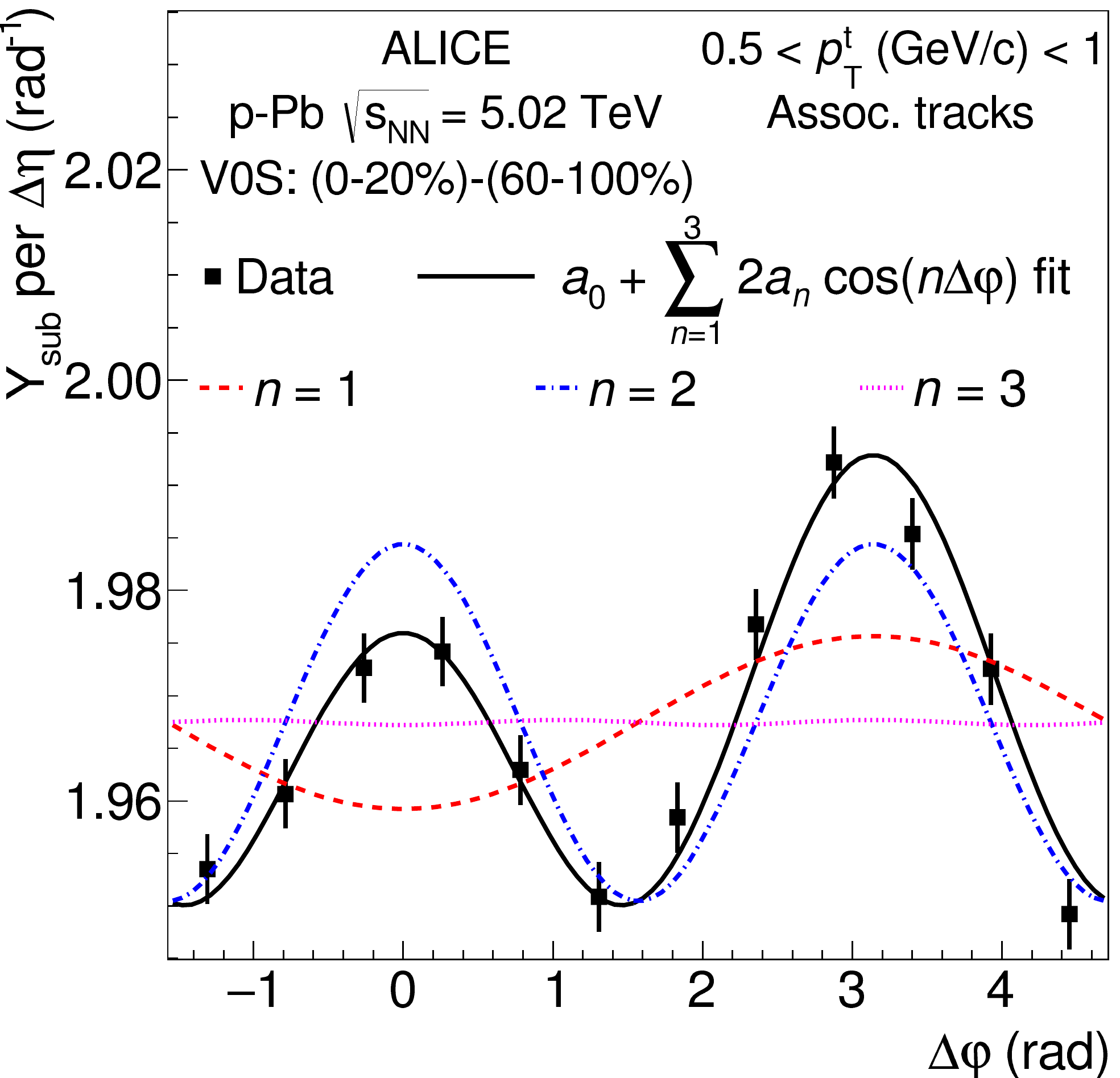}
\includegraphics[width=0.32\textwidth]{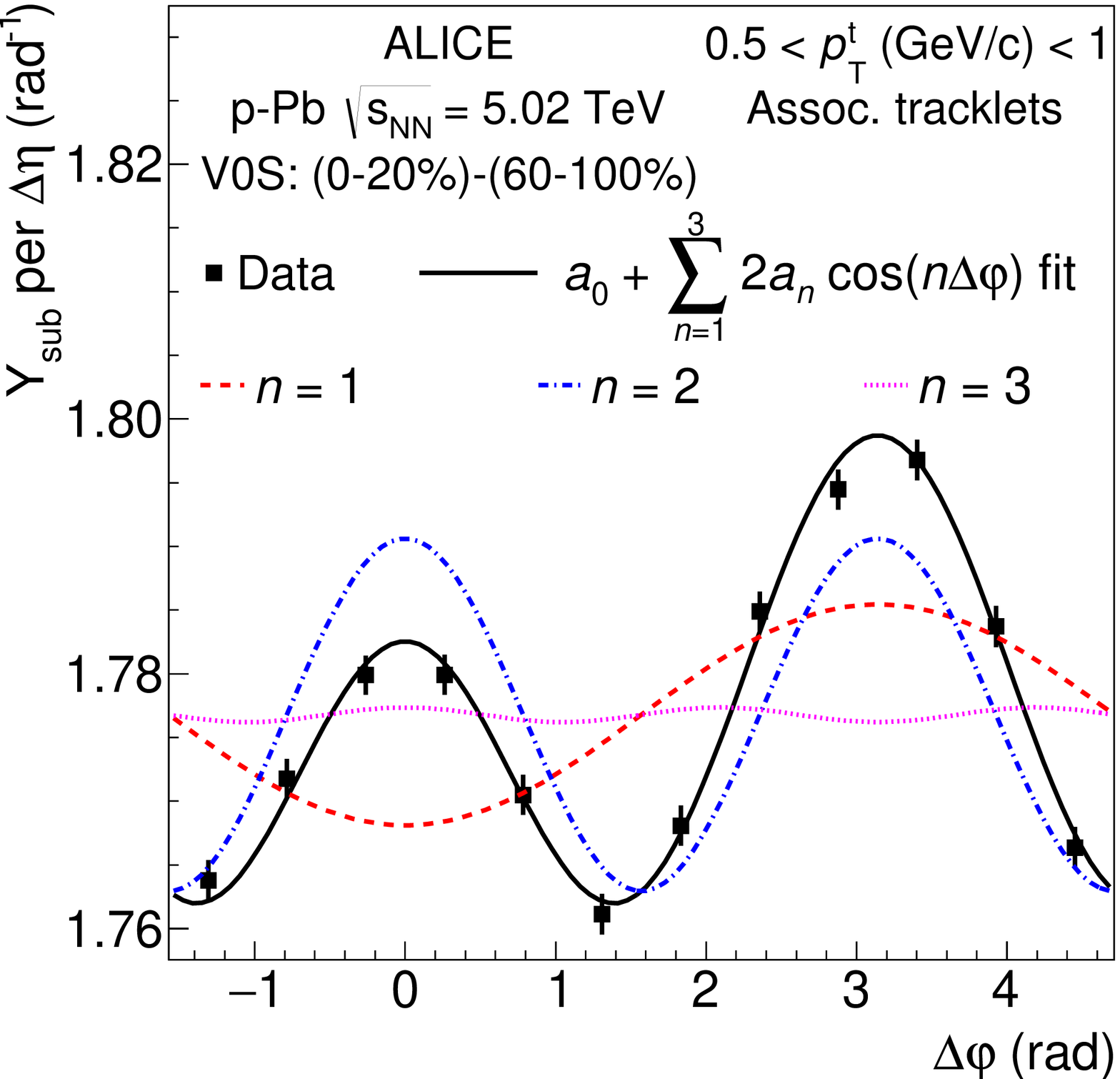}
\includegraphics[width=0.32\textwidth]{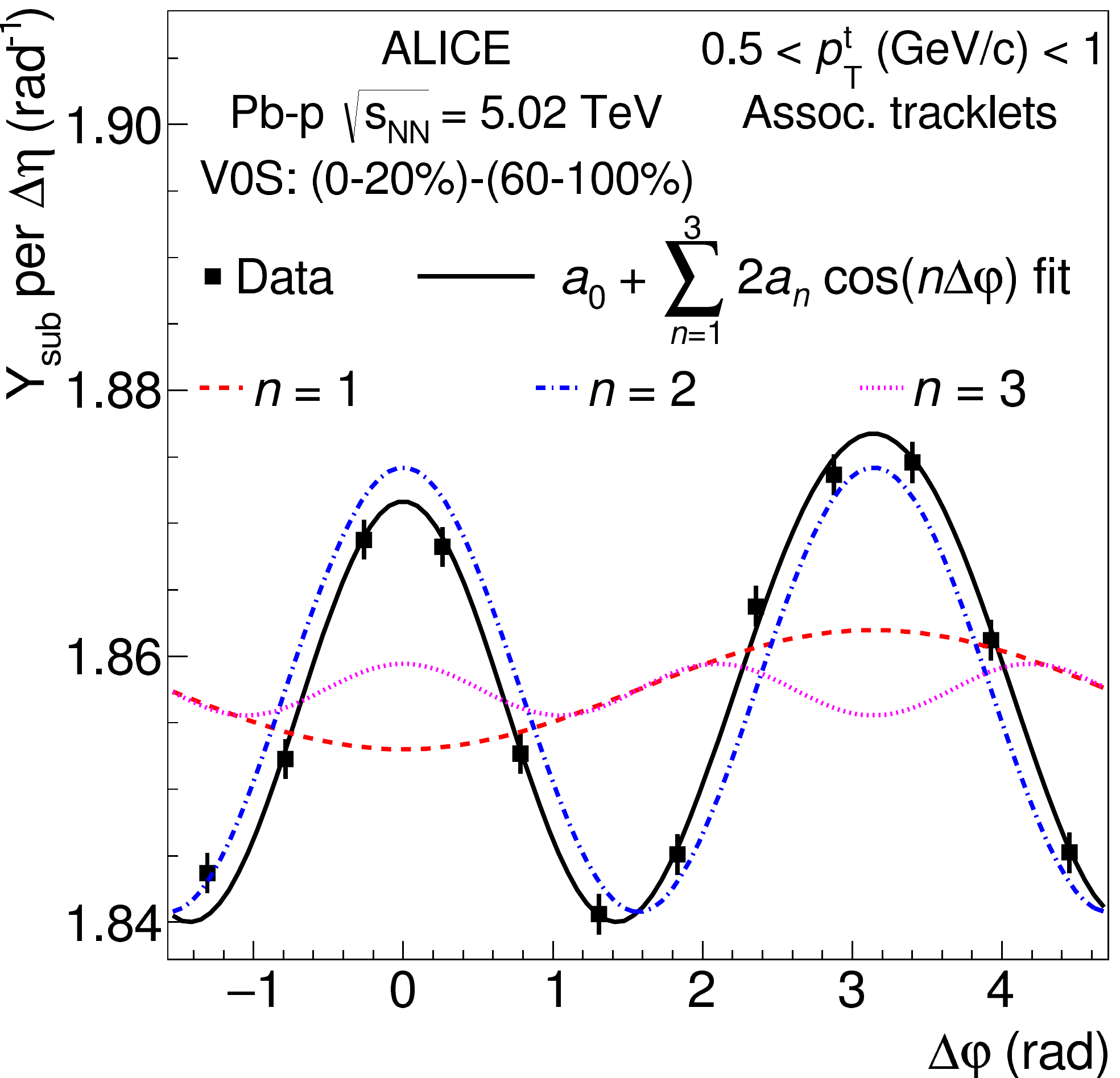}
\caption{\label{fig:corr_after_sub}
 Top panels: Associated yield per trigger particle as a function of $\Dphi$ and $\Deta$ for muon-track correlations in \pPb\ (left) 
 and muon-tracklet correlations in \pPb\ (centre) and \Pbp\ (right) collisions for the 0--20\% event class, where the corresponding correlation 
 from the 60--100\% event class has been subtracted. Statistical uncertainties are not shown.
 The trigger particle (muon) range is $0.5<\ptt<1$~\gevc, 
 the associated particle intervals are $0.5<\pta<4.0$~\gevc\ for tracks and \mbox{$0<\Delta\varphi_{\,\rm h}<5$}~mrad for tracklets.
 Bottom panels: The same as above projected onto $\Dphi$. The lines indicate the fit to the data and the first harmonic contributions 
 as explained in the text.
}
\end{figure}

In order to isolate long-range correlations, we apply the same subtraction method as in previous measurements~\cite{Abelev:2012ola,Abelev:2013wsa}. 
Jet-associated yields have only a weak multiplicity dependence~\cite{Abelev:2014mva}, 
thus the subtraction of the low-multiplicity event class removes most of the jet-like correlations.
The per-trigger yield of the 60--100\% event class is subtracted from that in the 0--20\% event class, 
and the result is presented~(labelled as $Y_{\rm sub}$) in the top panels of \Fig{fig:corr_after_sub}. 
After subtraction, two similar ridges on the near and on the away side are clearly visible.

The magnitude of the contributing long-range amplitudes is quantified by extracting the Fourier coefficients from the $\Dphi$ projection 
of the per-trigger yield distribution, after the subtraction of the low-multiplicity class, as shown in the lower panels of \Fig{fig:corr_after_sub}.
In order to reduce the statistical fluctuations at the edges of the per-trigger yield distribution, the \Dphi\ projection is obtained from a 
first-order polynomial fit along \Deta\ for each \Dphi\ interval. 
In the \pPb\ cases, the near- and away-side amplitudes are quite different, while in the \Pbp\ case the amplitudes on the near and away side are similar. 
The difference in the amplitudes of the near- and away-side ridge, which may be due to a residual jet contribution in the subtracted distribution, 
is taken into account in the systematic error evaluation, as explained in \Sect{sec:systematics}.

The Fourier coefficients are then obtained by fitting $Y_{\rm sub}$ with
\begin{equation}
a_0 + 2\,a_1 \cos (\Dphi) + 2\,a_2 \cos (2\Dphi) + 2\,a_3 \cos (3\Dphi)\,,
\label{fitfunction1}
\end{equation}
leading to $\chi^2/{\rm NDF}$ values typically below $1.5$.
The relative modulation is given by $V_{n\Delta}\{{\rm 2PC,sub}\} = \frac{a_n}{a_0 + b}$, where $b$ is the baseline of the low-multiplicity class~(60--100\%) 
estimated from the integral of the per-trigger yield around the minimum. 
Assuming that the two-particle Fourier coefficient factorizes into a product of trigger and associate single-particle $v_{2}$~\cite{Aad:2014lta},
the \vnsub\ coefficients for particles reconstructed in the \muonarm\ are then obtained as
\begin{eqnarray}
\vnsub = V_{n\Delta}\{{\rm 2PC,sub}\} / \sqrt{V_{n\Delta}^{\rm c}\{{\rm 2PC,sub}\}},
\label{vn}
\end{eqnarray}
where $V_{n\Delta}^{\rm c}\{{\rm 2PC,sub}\}$ is measured by correlating only central barrel tracks (or tracklets) with each other~(essentially repeating
the analysis as in \Ref{Abelev:2012ola}). 

In this Letter, $\vtwo$ values for muons in the acceptance of the \muonarm\ are reported.
Weak decays and scattering in the absorber of the \muonarm\ can cause the kinematics of reconstructed muons to deviate
from those of their parent particles, and can influence the reconstructed $v_{2}$,
especially in case $v_{2,\text{parent}}$ has a strong $\pt$ dependence.
Since we cannot correct the measured $v_{2}$ for the species-dependent inefficiencies induced by the absorber,
we denote the resulting coefficients by $\vtwomu$ to indicate that the result holds for decay muons measured in the \muonarm.

% $Id: fmc_systematics.tex 1312 2015-11-08 18:40:33Z loizides $
%%%%%%%%%%%%%%%%%%%%%%%%%%%%%%%%%%%%%%%%%%%%%%%%%%%%%%%%%%%%%%%%%%%%%%%%%%%%%%%%%%%%%%%%%%%%%%%%%%%%
\section{Systematic uncertainties}
\label{sec:systematics}
%%%%%%%%%%%%%%%%%%%%%%%%%%%%%%%%%%%%%%%%%%%%%%%%%%%%%%%%%%%%%%%%%%%%%%%%%%%%%%%%%%%%%%%%%%%%%%%%%%%%
The systematic uncertainty on $\vtwomu$ was estimated by varying the analysis procedure as described in this section. 
The uncertainty on the ratio between the $\vtwomu$ in \Pbp\ and \pPb\ collisions was obtained on the ratio 
itself, in order to properly treat the (anti-) correlated systematics between the \pPb\ and \Pbp\ data samples. 
A summary is given in \Tab{tab:syst}.

\begin{table}[t]
\centering
\begin{tabular}{l|c|c|c|c}
                   & Assoc. tracks  &  \multicolumn{3}{c}{Assoc. tracklets} \\
 Systematic effect & \pPb & \pPb & \Pbp & Ratio \\
\hline
 Acceptance (\zvtx\ dependence)            &3$-$4\%&0$-$5\%&0$-$3\%&0$-$1\%\\
 Remaining jet after subtraction           &4$-$10\%&5$-$14\%&1$-$2\%&3$-$15\%\\
 Remaining ridge in low-multiplicity class &1$-$4\%&1$-$6\%&0$-$2\%&2$-$8\%\\
 Calculation of $v_2$                      &0$-$1\%&0$-$1\%&1\%&0$-$2\%\\
 Resolution correction                     &1\%&0$-$1\%&0$-$1\%&0$-$2\%\\
\hline
Sum (added in quadrature)                  &7$-$11\%&6$-$14\%&2$-$4\%&5$-$17\%\\
\end{tabular}
\caption{Summary of main systematic uncertainties. The uncertainties usually depend on \pt\ and vary within the given ranges.}
\label{tab:syst}
\end{table}

%\subsection*{Acceptance and efficiency correction}
The acceptance of the ALICE central barrel depends on the position of \zvtx. %the primary vertex interaction. 
To study its influence on $\vtwomu$, the analysis was repeated using only events with a reconstructed primary vertex 
within $\pm$\unit[5]{cm} instead of $\pm$\unit[7]{cm} from the nominal interaction point.
The yield per trigger particle was not corrected for single track acceptance and efficiency of associated particles. 
Since $\vtwomu$ is a relative quantity, it is not expected to depend on the normalization. 
This was verified in the case of the muon-track analysis, where good agreement was found between the second-order Fourier coefficients obtained 
with and without single-track acceptance and efficiency corrections.
Hence, no additional uncertainty was considered.

%\subsection*{Remaining jet after subtraction}
As observed in previous analyses~\cite{Abelev:2012ola,Abelev:2013wsa}, the subtraction of the low-multiplicity class leads to a residual 
peak around $(\Deta,\Dphi) \approx (0,0)$, possibly due to a bias of the event selection on the jet fragmentation in low-multiplicity events~\cite{Abelev:2014mva}.
The pseudorapidity gap~\cite{Abelev:2012ola,Aad:2012gla} used to calculate $V_{n\Delta}^{c}$ was varied from $1.2$ to $1.0$ and to
$0.8$ in order to estimate the contribution of the residual near-side short-range correlations. 
Due to the large gap in pseudorapidity between the ALICE central barrel and the \muonarm, this contribution does not affect the forward-central correlation.
The effect of the bias introduced by the multiplicity selection was addressed on the away side by scaling the $60$--$100$\% multiplicity class. 
The scaling factor ($f$) is determined as the ratio between away-side yields in high- and low-multiplicity classes after the subtraction 
of the second-order Fourier component~\cite{Abelev:2014mva}. 
This procedure was applied in the calculation of both $V_{n\Delta}$ and $V_{n\Delta}^{\rm c}$. 
The scaling factors were found to be larger in the case of \pPb\ collisions~($f\leq 1.40$), compared to \Pbp~($f\leq 1.26$), and tend to
be lower for increasing $\pt$.
The difference with respect to the baseline results, for which no scaling ($f=1$) is applied, was taken as the systematic uncertainty.

%\subsection*{Remaining ridge in low-multiplicity class}
As previously reported~\cite{Abelev:2014mva}, the contribution of the long-range correlations to the measured yields is not significant 
in low-multiplicity events. 
Still, their potential influence was addressed by changing the multiplicity range from $60$--$100$\% to $70$--$100$\% for the low-multiplicity class.

%\subsection*{Calculation of $v_2$}
To test the stability of the fit,
the $v_2$ coefficient was calculated using a fit with only the first and the second Fourier components in \Eq{fitfunction1}. 
As another variation, the baseline $b$ was calculated from a fit of the per-trigger yield in the low-multiplicity class 
using a Gaussian to model the shape of the away-side ridge and a constant to estimate $b$. 
An equivalent approach, which makes use of the baseline of the high-multiplicity class $B$ in $V_{n\Delta}\{{\rm 2PC,sub}\} = a_n / B$, was also used,
where $B$ was estimated from the integral or from a parabolic fit of the correlation function around the minimum.
Finally, the $\Dphi$ projection was obtained from a weighted average instead of a first-order polynomial fit along $\Deta$ for each $\Dphi$ interval.

%\subsection*{Resolution correction}
The effect from the finite angular and momentum resolution of the \muonarm\ on $\vtwomu$ was evaluated from a dedicated MC study with the measured $v_2$ as input
distribution, and resulted in a small correction of below $2$\%. 
The associated uncertainty was evaluated by varying the input $v_2$ by $50$\% at the lowest and highest measured points. 

% $Id: fmc_results.tex 1342 2015-11-29 17:13:32Z loizides $

%%%%%%%%%%%%%%%%%%%%%%%%%%%%%%%%%%%%%%%%%%%%%%%%%%%%%%%%%%%%%%%%%%%%%%%%%%%%%%%%%%%%%%%%%%%%%%%%%%%%
\section{Results}
\label{sec:results}
\begin{figure}[t]
\centering
\includegraphics[width=0.49\textwidth]{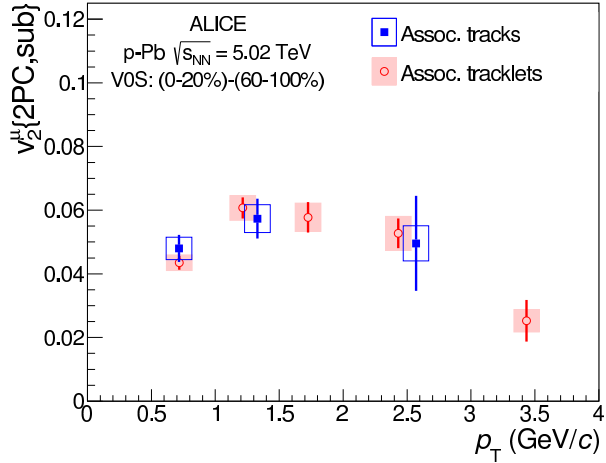}
\caption{\label{fig:v2comp}
 Comparison of \vtwomu\ for $-4<\eta<-2.5$ extracted from muon-track and muon-tracklet correlations in \pPb\ collisions at $\snn=5.02$~TeV.
}
\end{figure}

%%%%%%%%%%%%%%%%%%%%%%%%%%%%%%%%%%%%%%%%%%%%%%%%%%%%%%%%%%%%%%%%%%%%%%%%%%%%%%%%%%%%%%%%%%%%%%%%%%%%
The $\vtwomu$ coefficients were measured for muon tracks in the p-going direction (\pPb\ period) using both tracks and tracklets 
as associated central barrel particles, as described in \Sect{sec:analysis}.
The $\vtwomu$ coefficients obtained from the per-trigger yields of associated central barrel tracks
agree well with those of associated tracklets, as shown in~\Fig{fig:v2comp} as a function of muon $\pt$. 
Since the two measurements probe different ranges in associated particle $\pt$, the agreement is a consequence of
trigger and associate $v_{2}$ factorization~\cite{Aad:2014lta}. 
In addition, good agreement was found between the $\vtwomu$ obtained with different cuts on $\Delta\varphi_{\,\rm h}$ 
of associated tracklets~(inducing a change of average $\pt$ by about $20$\%). 

\begin{figure}[t]
\centering
\includegraphics[width=0.49\textwidth]{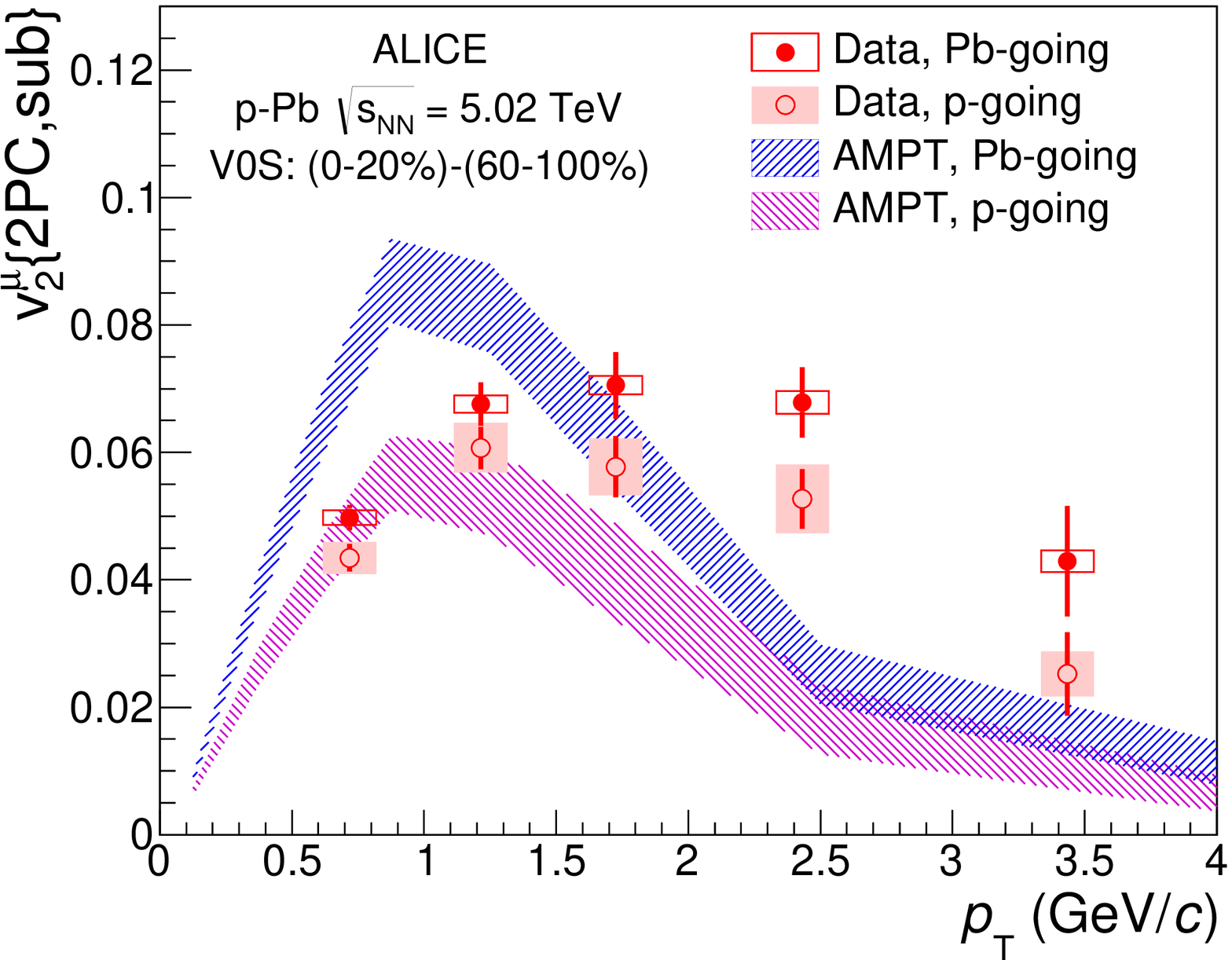}
\includegraphics[width=0.49\textwidth]{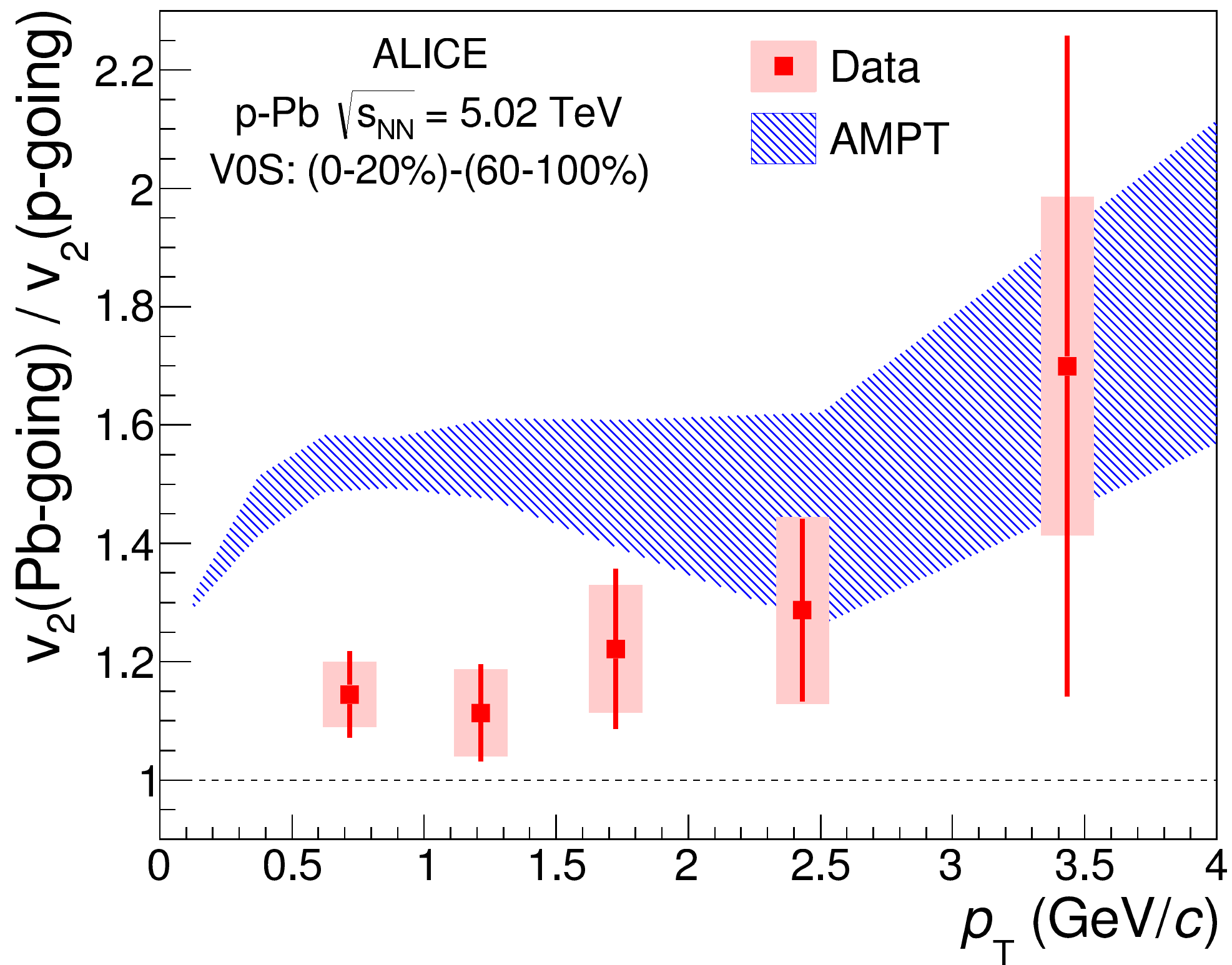}
\caption{\label{fig:v2bf}
 The \vtwomu\ coefficients from muon-tracklet correlations in p-going and Pb-going directions (left) and their ratio (right)
 for $-4<\eta<-2.5$ in \pPb\ collisions at $\snn=5.02$~TeV. The data are compared to model calculations from AMPT.
}
\end{figure}

The p-going and Pb-going \vtwomu\ coefficients obtained using muon-tracklet correlations for the two different beam 
configurations~(\pPb\ and \Pbp) are reported in the left panel of \Fig{fig:v2bf} as a function of muon $\pt$. 
The Pb-going \vtwomu\ (i.e.\ when the muon trigger particle travels in the same direction as the Pb nucleus) is observed 
to be larger than the p-going \vtwomu\ over the measured \pt\ range, but the two have a similar $\pt$-dependence.
To quantify the asymmetry, the Pb-going over p-going ratio\com{ (forward-over-backward ratio)} for the \vtwomu\ 
coefficients is reported in the right panel of \Fig{fig:v2bf} as a function of muon $\pt$. 
The ratio is found to be rather independent of $\pt$ given the statistical and systematic uncertainties of the measurement.
A constant fit to the ratio adding statistical and systematic uncertainties in quadrature gives $1.16\pm 0.06$ with a $\chi^2/{\rm NDF}=0.4$.
The analysis was also repeated using the energy deposited in the neutron ZDCs on the Pb-going side instead of the V0S amplitude for the event class definition. 
As discussed in detail in~\cite{Adam:2014qja}, the correlation between forward energy measured in the ZDCs and particle density at central rapidities 
is weak in \pPb\ collisions.
Therefore, event classes defined as fixed fractions of the signal distribution in the ZDCs select different events, with different mean particle 
multiplicity at midrapidity, than the samples selected with the same fractions in the \VZERO\ detector.
Still, the \vtwomu\ values were measured to be similar, within 25\% of those extracted with V0S estimator. 
In addition, the asymmetry between Pb- and p-going \vtwomu\ was found to persist with similar shape and magnitude. 
The observed asymmetry may result from decorrelations of event planes at different rapidity~\cite{Khachatryan:2015oea}.


The data in \Fig{fig:v2bf} can not be readily compared with existing predictions~\cite{Bozek:2015swa} for a 3+1~dimensional, viscous hydrodynamical 
model~\cite{Bozek:2011if} and the AMPT model with the string-melting mechanism enabled~\cite{Lin:2004en}. 
The model calculations were performed without taking into account the effect of the muon absorber, and represent the $v_2$ of primary particles,
while as discussed in \Sect{sec:selection} the measured $\vtwomu$ coefficients are reported for decay muons.
Depending on particle composition and on the $\pt$-dependence of the parent particle $v_2$ distribution, 
the difference between primary particle $v_2$ and decay muon $v_2$ can be quite large.
For example, at $1$~\gevc, assuming the $v_2$ of the parent particles rises with $\pt$ like at mid-rapidity~\cite{Abelev:2013wsa}, 
the measured $\vtwomu$ for muons originating from decays of pions~(kaons) would be $\approx$20~(40)\% larger than that of the parent pions~(kaons). 

Instead, in \Fig{fig:v2bf} we show a comparison of the data with AMPT model calculations performed with the same parameters as in \cite{Bozek:2015swa}.
These calculations were performed at generator level, decaying primary particles into muons using the PYTHIA decayer~\cite{Sjostrand:2006za}.
The effects of the muon absorber were included by applying the $\pt$ and $\eta$ dependent 
relative efficiencies provided in the right panel of \Fig{fig:mumothers}.
Event characterization was done by mimicking the V0S criteria at particle level, i.e.\ by counting charged particles in $2.8<\eta<3.9$ 
and $-3.7<\eta<-2.7$.
The $v_2$ values were obtained separately for muons decaying from pions, kaons and heavy-flavor hadrons, 
and otherwise performing the analysis in the same way as in data. %, i.e.\ by correlating muons with charged particles in the respective acceptances,
%and by subtracting the resulting correlations in low-multiplicity from high-multiplicity events. 
We found the $v_2$ for HF muons to be consistent with zero within the generated statistics (5M events with a HF muon in the acceptance
of the \muonarm\ for each period). 
Hence, for the inclusive $v_2$, which is obtained by weighting the calculated $v_2$ with the relative yields in each decay channel, 
the $v_2$ for HF muons has been set to zero to reduce statistical fluctuations.
In AMPT the factor $f$ used to scale low-multiplicity class to eliminate the remaining jet contribution after subtraction, reaches values much larger 
than in the data, up to $f=2$. 
Applying the scaling reduces the extracted $v_2$ and consequently this choice constitutes the lower~(upper) bound of the shaded area in \Fig{fig:v2bf} left (right),
while the opposite bounds correspond to $f=1$ (as used for the baseline result in the data). 

As shown in the left panel of \Fig{fig:v2bf}, below $\pt<1.5$~\gevc, where the inclusive muon yield is expected to be dominated by weak decays of pions and kaons, 
the calculation produces qualitatively similar trends as observed in the data. 
However, quantitatively a different $\pt$ and $\eta$ dependence is found, visible in particular in the right panel of \Fig{fig:v2bf}.
At $\pt>2$~\gevc, where the inclusive muon yield is dominated by heavy-flavor decays, the data may support a finite value for the $v_2$ of
HF muons, or a drastically different composition of the parent distribution or their $v_2$ values in AMPT compared to data.
Indeed, comparing predictions of AMPT to pion, kaon and D-meson yields measured at midrapidity~\cite{Abelev:2013haa,Abelev:2014hha}, muons 
from heavy-flavor decays would be underestimated by a factor $3$--$5$ relative to pion and kaon decays assuming the same discrepancy
between model and data at forward rapidity. %increasing the support for a finite value for HF muon $v_2$.
A finite value for HF muon $v_2$ would be consistent with the emergence of radial flow in heavy-flavor meson spectra as predicted in~\cite{Sickles:2013yna},
and has been recently measured in \PbPb\ collisions at $\snn=2.76$~TeV~\cite{Adam:2015pga}.

% $Id: fmc_summary.tex 1312 2015-11-08 18:40:33Z loizides $

%%%%%%%%%%%%%%%%%%%%%%%%%%%%%%%%%%%%%%%%%%%%%%%%%%%%%%%%%%%%%%%%%%%%%%%%%%%%%%%%%%%%%%%%%%%%%%%%%%%%
\section{Summary}
\label{sec:summary}
%%%%%%%%%%%%%%%%%%%%%%%%%%%%%%%%%%%%%%%%%%%%%%%%%%%%%%%%%%%%%%%%%%%%%%%%%%%%%%%%%%%%%%%%%%%%%%%%%%%%
Two-particle angular correlations between trigger particles in the forward pseudorapidity range $2.5 < |\eta| < 4.0$ 
and associated particles in the central range $|\eta| < 1.0$ measured by ALICE are reported in \pPb\ collisions 
at a nucleon--nucleon centre-of-mass energy of \unit[5.02]{TeV}.
The trigger particles are inclusive muons and the associated particles are charged particles,
reconstructed by the muon spectrometer and central barrel tracking detectors, respectively.
The composition of parent particles for the measured muons is expected to vary as a function of $\pt$ (\Fig{fig:mumothers}).
A near-side ridge is observed in high-multiplicity events~(\Fig{fig:corr_before_sub}).
After subtraction of jet-like correlations measured in low-multiplicity events, the double-ridge structure, 
previously discovered in two-particle angular correlations at midrapidity, is found to persist even in the pseudorapidity 
ranges studied here~(\Fig{fig:corr_after_sub}).
The second-order Fourier coefficients for muon tracks are determined assuming factorization of the
Fourier coefficients at central and forward rapidity.
The measurement in \pPb\ collisions was performed in two different ways, using tracks or tracklets for particles
at $|\eta| < 1.0$, yielding consistent results~(\Fig{fig:v2comp}).
The second-order Fourier coefficients for muons in high-multiplicity events were found to have a similar transverse 
momentum dependence in the p-going (\pPb) and Pb-going (\Pbp) configurations, with the Pb-going coefficients 
larger by $16\pm6$\%, rather independent of $\pt$ within the uncertainties of the measurement~(\Fig{fig:v2bf}).
The results were compared with calculations using the AMPT model incorporating the effects of the muon absorber, 
showing a different $\pt$ and $\eta$ dependence than observed in the data.
Above $2$~\gevc, the results are sensitive to the $v_2$ of heavy-flavor decay muons.
Forthcoming model calculations should apply the relative efficiencies for muon decays from pion and kaons~(provided 
in \Fig{fig:mumothers}) at generator level for detailed comparison with our data.
Further measurements (e.g.\ of heavy-flavor muon yields or charged-particle $v_2$ at forward rapidity) will be needed 
to reduce the ambiguity between between muon parent particle composition and their $v_2$.

%==========================================================%
%======================ACK+BIBLIO==========================%
%==========================================================%
\ifpreprint
\iffull
\newenvironment{acknowledgement}{\relax}{\relax}
\begin{acknowledgement}
\section*{Acknowledgements}
% $Id: acknowledgements.tex 2098 2015-04-24 15:54:16Z loizides $
% Version: Jan 2015
We thank P.\ Bozek, A.\ Bzdak, and G.-L.\ Ma for fruitful discussions concerning their calculations~\cite{Bozek:2015swa}.

The ALICE Collaboration would like to thank all its engineers and technicians for their invaluable contributions to the construction of the experiment and the CERN accelerator teams for the outstanding performance of the LHC complex.
The ALICE Collaboration gratefully acknowledges the resources and support provided by all Grid centres and the Worldwide LHC Computing Grid (WLCG) collaboration.
The ALICE Collaboration acknowledges the following funding agencies for their support in building and
running the ALICE detector:
State Committee of Science,  World Federation of Scientists (WFS)
and Swiss Fonds Kidagan, Armenia,
Conselho Nacional de Desenvolvimento Cient\'{\i}fico e Tecnol\'{o}gico (CNPq), Financiadora de Estudos e Projetos (FINEP),
Funda\c{c}\~{a}o de Amparo \`{a} Pesquisa do Estado de S\~{a}o Paulo (FAPESP);
National Natural Science Foundation of China (NSFC), the Chinese Ministry of Education (CMOE)
and the Ministry of Science and Technology of China (MSTC);
Ministry of Education and Youth of the Czech Republic;
Danish Natural Science Research Council, the Carlsberg Foundation and the Danish National Research Foundation;
The European Research Council under the European Community's Seventh Framework Programme;
Helsinki Institute of Physics and the Academy of Finland;
French CNRS-IN2P3, the `Region Pays de Loire', `Region Alsace', `Region Auvergne' and CEA, France;
German Bundesministerium fur Bildung, Wissenschaft, Forschung und Technologie (BMBF) and the Helmholtz Association;
General Secretariat for Research and Technology, Ministry of
Development, Greece;
Hungarian Orszagos Tudomanyos Kutatasi Alappgrammok (OTKA) and National Office for Research and Technology (NKTH);
Department of Atomic Energy and Department of Science and Technology of the Government of India;
Istituto Nazionale di Fisica Nucleare (INFN) and Centro Fermi -
Museo Storico della Fisica e Centro Studi e Ricerche "Enrico
Fermi", Italy;
MEXT Grant-in-Aid for Specially Promoted Research, Ja\-pan;
Joint Institute for Nuclear Research, Dubna;
National Research Foundation of Korea (NRF);
Consejo Nacional de Cienca y Tecnologia (CONACYT), Direccion General de Asuntos del Personal Academico(DGAPA), M\'{e}xico, Amerique Latine Formation academique - European Commission~(ALFA-EC) and the EPLANET Program~(European Particle Physics Latin American Network);
Stichting voor Fundamenteel Onderzoek der Materie (FOM) and the Nederlandse Organisatie voor Wetenschappelijk Onderzoek (NWO), Netherlands;
Research Council of Norway (NFR);
National Science Centre, Poland;
Ministry of National Education/Institute for Atomic Physics and National Council of Scientific Research in Higher Education~(CNCSI-UEFISCDI), Romania;
Ministry of Education and Science of Russian Federation, Russian
Academy of Sciences, Russian Federal Agency of Atomic Energy,
Russian Federal Agency for Science and Innovations and The Russian
Foundation for Basic Research;
Ministry of Education of Slovakia;
Department of Science and Technology, South Africa;
Centro de Investigaciones Energeticas, Medioambientales y Tecnologicas (CIEMAT), E-Infrastructure shared between Europe and Latin America (EELA), Ministerio de Econom\'{i}a y Competitividad (MINECO) of Spain, Xunta de Galicia (Conseller\'{\i}a de Educaci\'{o}n),
Centro de Aplicaciones Tecnológicas y Desarrollo Nuclear (CEA\-DEN), Cubaenerg\'{\i}a, Cuba, and IAEA (International Atomic Energy Agency);
Swedish Research Council (VR) and Knut $\&$ Alice Wallenberg
Foundation (KAW);
Ukraine Ministry of Education and Science;
United Kingdom Science and Technology Facilities Council (STFC);
The United States Department of Energy, the United States National
Science Foundation, the State of Texas, and the State of Ohio;
Ministry of Science, Education and Sports of Croatia and  Unity through Knowledge Fund, Croatia.
Council of Scientific and Industrial Research (CSIR), New Delhi, India
        %%%%%%% get the latest version before submitting
\end{acknowledgement}
\ifbibtex
\bibliographystyle{utphys}
\bibliography{biblio}{}
\else
\input{refpreprint.tex}
\fi
\newpage
\appendix
\section{The ALICE Collaboration}
\label{app:collab}

% Collaboration: CERN-LHC-ALICE
% Generation Date is 2015/Jun/18

% How to use:
%%%%%%%%% appendix with author list
%\appendix
%\section{The ALICE Collaboration}
%\label{app:collab}
%\input{authors-list.tex}  %%%%%%% get the latest version before submitting

\begingroup
\small
\begin{flushleft}
J.~Adam\Irefn{org40}\And
D.~Adamov\'{a}\Irefn{org83}\And
M.M.~Aggarwal\Irefn{org87}\And
G.~Aglieri Rinella\Irefn{org36}\And
M.~Agnello\Irefn{org111}\And
N.~Agrawal\Irefn{org48}\And
Z.~Ahammed\Irefn{org132}\And
S.U.~Ahn\Irefn{org68}\And
I.~Aimo\Irefn{org94}\textsuperscript{,}\Irefn{org111}\And
S.~Aiola\Irefn{org136}
% ...
\renewcommand\labelenumi{\textsuperscript{\theenumi}~}

\section*{Affiliation notes}
\renewcommand\theenumi{\roman{enumi}}
\begin{Authlist}
\item \Adef{0}Deceased
\item \Adef{idp3813520}{Also at: M.V. Lomonosov Moscow State University, D.V. Skobeltsyn Institute of Nuclear, Physics, Moscow, Russia}
\item \Adef{idp5954704}{Also at: University of Kansas, Lawrence, Kansas, United States}
\end{Authlist}

\section*{Collaboration Institutes}
\renewcommand\theenumi{\arabic{enumi}~}
\begin{Authlist}

\item \Idef{org1}A.I. Alikhanyan National Science Laboratory (Yerevan Physics Institute) Foundation, Yerevan, Armenia
\item \Idef{org2}Benem\'{e}rita Universidad Aut\'{o}noma de Puebla, Puebla, Mexico
\item \Idef{org3}Bogolyubov Institute for Theoretical Physics, Kiev, Ukraine
\item \Idef{org4}Bose Institute, Department of Physics and Centre for Astroparticle Physics and Space Science (CAPSS), Kolkata, India
\item \Idef{org5}Budker Institute for Nuclear Physics, Novosibirsk, Russia
\item \Idef{org6}California Polytechnic State University, San Luis Obispo, California, United States
\item \Idef{org7}Central China Normal University, Wuhan, China
\item \Idef{org8}Centre de Calcul de l'IN2P3, Villeurbanne, France
\item \Idef{org9}Centro de Aplicaciones Tecnol\'{o}gicas y Desarrollo Nuclear (CEADEN), Havana, Cuba
\item \Idef{org10}Centro de Investigaciones Energ\'{e}ticas Medioambientales y Tecnol\'{o}gicas (CIEMAT), Madrid, Spain
\item \Idef{org11}Centro de Investigaci\'{o}n y de Estudios Avanzados (CINVESTAV), Mexico City and M\'{e}rida, Mexico
\item \Idef{org12}Centro Fermi - Museo Storico della Fisica e Centro Studi e Ricerche ``Enrico Fermi'', Rome, Italy
\item \Idef{org13}Chicago State University, Chicago, Illinois, USA
\item \Idef{org14}China Institute of Atomic Energy, Beijing, China
\item \Idef{org15}Commissariat \`{a} l'Energie Atomique, IRFU, Saclay, France
\item \Idef{org16}COMSATS Institute of Information Technology (CIIT), Islamabad, Pakistan
\item \Idef{org17}Departamento de F\'{\i}sica de Part\'{\i}culas and IGFAE, Universidad de Santiago de Compostela, Santiago de Compostela, Spain
\item \Idef{org18}Department of Physics and Technology, University of Bergen, Bergen, Norway
\item \Idef{org19}Department of Physics, Aligarh Muslim University, Aligarh, India
\item \Idef{org20}Department of Physics, Ohio State University, Columbus, Ohio, United States
\item \Idef{org21}Department of Physics, Sejong University, Seoul, South Korea
\item \Idef{org22}Department of Physics, University of Oslo, Oslo, Norway
\item \Idef{org23}Dipartimento di Elettrotecnica ed Elettronica del Politecnico, Bari, Italy
\item \Idef{org24}Dipartimento di Fisica dell'Universit\`{a} 'La Sapienza' and Sezione INFN Rome, Italy
\item \Idef{org25}Dipartimento di Fisica dell'Universit\`{a} and Sezione INFN, Cagliari, Italy
\item \Idef{org26}Dipartimento di Fisica dell'Universit\`{a} and Sezione INFN, Trieste, Italy
\item \Idef{org27}Dipartimento di Fisica dell'Universit\`{a} and Sezione INFN, Turin, Italy
\item \Idef{org28}Dipartimento di Fisica e Astronomia dell'Universit\`{a} and Sezione INFN, Bologna, Italy
\item \Idef{org29}Dipartimento di Fisica e Astronomia dell'Universit\`{a} and Sezione INFN, Catania, Italy
\item \Idef{org30}Dipartimento di Fisica e Astronomia dell'Universit\`{a} and Sezione INFN, Padova, Italy
\item \Idef{org31}Dipartimento di Fisica `E.R.~Caianiello' dell'Universit\`{a} and Gruppo Collegato INFN, Salerno, Italy
\item \Idef{org32}Dipartimento di Scienze e Innovazione Tecnologica dell'Universit\`{a} del  Piemonte Orientale and Gruppo Collegato INFN, Alessandria, Italy
\item \Idef{org33}Dipartimento Interateneo di Fisica `M.~Merlin' and Sezione INFN, Bari, Italy
\item \Idef{org34}Division of Experimental High Energy Physics, University of Lund, Lund, Sweden
\item \Idef{org35}Eberhard Karls Universit\"{a}t T\"{u}bingen, T\"{u}bingen, Germany
\item \Idef{org36}European Organization for Nuclear Research (CERN), Geneva, Switzerland
\item \Idef{org37}Excellence Cluster Universe, Technische Universit\"{a}t M\"{u}nchen, Munich, Germany
\item \Idef{org38}Faculty of Engineering, Bergen University College, Bergen, Norway
\item \Idef{org39}Faculty of Mathematics, Physics and Informatics, Comenius University, Bratislava, Slovakia
\item \Idef{org40}Faculty of Nuclear Sciences and Physical Engineering, Czech Technical University in Prague, Prague, Czech Republic
\item \Idef{org41}Faculty of Science, P.J.~\v{S}af\'{a}rik University, Ko\v{s}ice, Slovakia
\item \Idef{org42}Faculty of Technology, Buskerud and Vestfold University College, Vestfold, Norway
\item \Idef{org43}Frankfurt Institute for Advanced Studies, Johann Wolfgang Goethe-Universit\"{a}t Frankfurt, Frankfurt, Germany
\item \Idef{org44}Gangneung-Wonju National University, Gangneung, South Korea
\item \Idef{org45}Gauhati University, Department of Physics, Guwahati, India
\item \Idef{org46}Helsinki Institute of Physics (HIP), Helsinki, Finland
\item \Idef{org47}Hiroshima University, Hiroshima, Japan
\item \Idef{org48}Indian Institute of Technology Bombay (IIT), Mumbai, India
\item \Idef{org49}Indian Institute of Technology Indore, Indore (IITI), India
\item \Idef{org50}Inha University, Incheon, South Korea
\item \Idef{org51}Institut de Physique Nucl\'eaire d'Orsay (IPNO), Universit\'e Paris-Sud, CNRS-IN2P3, Orsay, France
\item \Idef{org52}Institut f\"{u}r Informatik, Johann Wolfgang Goethe-Universit\"{a}t Frankfurt, Frankfurt, Germany
\item \Idef{org53}Institut f\"{u}r Kernphysik, Johann Wolfgang Goethe-Universit\"{a}t Frankfurt, Frankfurt, Germany
\item \Idef{org54}Institut f\"{u}r Kernphysik, Westf\"{a}lische Wilhelms-Universit\"{a}t M\"{u}nster, M\"{u}nster, Germany
\item \Idef{org55}Institut Pluridisciplinaire Hubert Curien (IPHC), Universit\'{e} de Strasbourg, CNRS-IN2P3, Strasbourg, France
\item \Idef{org56}Institute for Nuclear Research, Academy of Sciences, Moscow, Russia
\item \Idef{org57}Institute for Subatomic Physics of Utrecht University, Utrecht, Netherlands
\item \Idef{org58}Institute for Theoretical and Experimental Physics, Moscow, Russia
\item \Idef{org59}Institute of Experimental Physics, Slovak Academy of Sciences, Ko\v{s}ice, Slovakia
\item \Idef{org60}Institute of Physics, Academy of Sciences of the Czech Republic, Prague, Czech Republic
\item \Idef{org61}Institute of Physics, Bhubaneswar, India
\item \Idef{org62}Institute of Space Science (ISS), Bucharest, Romania
\item \Idef{org63}Instituto de Ciencias Nucleares, Universidad Nacional Aut\'{o}noma de M\'{e}xico, Mexico City, Mexico
\item \Idef{org64}Instituto de F\'{\i}sica, Universidad Nacional Aut\'{o}noma de M\'{e}xico, Mexico City, Mexico
\item \Idef{org65}iThemba LABS, National Research Foundation, Somerset West, South Africa
\item \Idef{org66}Joint Institute for Nuclear Research (JINR), Dubna, Russia
\item \Idef{org67}Konkuk University, Seoul, South Korea
\item \Idef{org68}Korea Institute of Science and Technology Information, Daejeon, South Korea
\item \Idef{org69}KTO Karatay University, Konya, Turkey
\item \Idef{org70}Laboratoire de Physique Corpusculaire (LPC), Clermont Universit\'{e}, Universit\'{e} Blaise Pascal, CNRS--IN2P3, Clermont-Ferrand, France
\item \Idef{org71}Laboratoire de Physique Subatomique et de Cosmologie, Universit\'{e} Grenoble-Alpes, CNRS-IN2P3, Grenoble, France
\item \Idef{org72}Laboratori Nazionali di Frascati, INFN, Frascati, Italy
\item \Idef{org73}Laboratori Nazionali di Legnaro, INFN, Legnaro, Italy
\item \Idef{org74}Lawrence Berkeley National Laboratory, Berkeley, California, United States
\item \Idef{org75}Lawrence Livermore National Laboratory, Livermore, California, United States
\item \Idef{org76}Moscow Engineering Physics Institute, Moscow, Russia
\item \Idef{org77}National Centre for Nuclear Studies, Warsaw, Poland
\item \Idef{org78}National Institute for Physics and Nuclear Engineering, Bucharest, Romania
\item \Idef{org79}National Institute of Science Education and Research, Bhubaneswar, India
\item \Idef{org80}Niels Bohr Institute, University of Copenhagen, Copenhagen, Denmark
\item \Idef{org81}Nikhef, Nationaal instituut voor subatomaire fysica, Amsterdam, Netherlands
\item \Idef{org82}Nuclear Physics Group, STFC Daresbury Laboratory, Daresbury, United Kingdom
\item \Idef{org83}Nuclear Physics Institute, Academy of Sciences of the Czech Republic, \v{R}e\v{z} u Prahy, Czech Republic
\item \Idef{org84}Oak Ridge National Laboratory, Oak Ridge, Tennessee, United States
\item \Idef{org85}Petersburg Nuclear Physics Institute, Gatchina, Russia
\item \Idef{org86}Physics Department, Creighton University, Omaha, Nebraska, United States
\item \Idef{org87}Physics Department, Panjab University, Chandigarh, India
\item \Idef{org88}Physics Department, University of Athens, Athens, Greece
\item \Idef{org89}Physics Department, University of Cape Town, Cape Town, South Africa
\item \Idef{org90}Physics Department, University of Jammu, Jammu, India
\item \Idef{org91}Physics Department, University of Rajasthan, Jaipur, India
\item \Idef{org92}Physik Department, Technische Universit\"{a}t M\"{u}nchen, Munich, Germany
\item \Idef{org93}Physikalisches Institut, Ruprecht-Karls-Universit\"{a}t Heidelberg, Heidelberg, Germany
\item \Idef{org94}Politecnico di Torino, Turin, Italy
\item \Idef{org95}Purdue University, West Lafayette, Indiana, United States
\item \Idef{org96}Pusan National University, Pusan, South Korea
\item \Idef{org97}Research Division and ExtreMe Matter Institute EMMI, GSI Helmholtzzentrum f\"ur Schwerionenforschung, Darmstadt, Germany
\item \Idef{org98}Rudjer Bo\v{s}kovi\'{c} Institute, Zagreb, Croatia
\item \Idef{org99}Russian Federal Nuclear Center (VNIIEF), Sarov, Russia
\item \Idef{org100}Russian Research Centre Kurchatov Institute, Moscow, Russia
\item \Idef{org101}Saha Institute of Nuclear Physics, Kolkata, India
\item \Idef{org102}School of Physics and Astronomy, University of Birmingham, Birmingham, United Kingdom
\item \Idef{org103}Secci\'{o}n F\'{\i}sica, Departamento de Ciencias, Pontificia Universidad Cat\'{o}lica del Per\'{u}, Lima, Peru
\item \Idef{org104}Sezione INFN, Bari, Italy
\item \Idef{org105}Sezione INFN, Bologna, Italy
\item \Idef{org106}Sezione INFN, Cagliari, Italy
\item \Idef{org107}Sezione INFN, Catania, Italy
\item \Idef{org108}Sezione INFN, Padova, Italy
\item \Idef{org109}Sezione INFN, Rome, Italy
\item \Idef{org110}Sezione INFN, Trieste, Italy
\item \Idef{org111}Sezione INFN, Turin, Italy
\item \Idef{org112}SSC IHEP of NRC Kurchatov institute, Protvino, Russia
\item \Idef{org113}SUBATECH, Ecole des Mines de Nantes, Universit\'{e} de Nantes, CNRS-IN2P3, Nantes, France
\item \Idef{org114}Suranaree University of Technology, Nakhon Ratchasima, Thailand
\item \Idef{org115}Technical University of Ko\v{s}ice, Ko\v{s}ice, Slovakia
\item \Idef{org116}Technical University of Split FESB, Split, Croatia
\item \Idef{org117}The Henryk Niewodniczanski Institute of Nuclear Physics, Polish Academy of Sciences, Cracow, Poland
\item \Idef{org118}The University of Texas at Austin, Physics Department, Austin, Texas, USA
\item \Idef{org119}Universidad Aut\'{o}noma de Sinaloa, Culiac\'{a}n, Mexico
\item \Idef{org120}Universidade de S\~{a}o Paulo (USP), S\~{a}o Paulo, Brazil
\item \Idef{org121}Universidade Estadual de Campinas (UNICAMP), Campinas, Brazil
\item \Idef{org122}University of Houston, Houston, Texas, United States
\item \Idef{org123}University of Jyv\"{a}skyl\"{a}, Jyv\"{a}skyl\"{a}, Finland
\item \Idef{org124}University of Liverpool, Liverpool, United Kingdom
\item \Idef{org125}University of Tennessee, Knoxville, Tennessee, United States
\item \Idef{org126}University of the Witwatersrand, Johannesburg, South Africa
\item \Idef{org127}University of Tokyo, Tokyo, Japan
\item \Idef{org128}University of Tsukuba, Tsukuba, Japan
\item \Idef{org129}University of Zagreb, Zagreb, Croatia
\item \Idef{org130}Universit\'{e} de Lyon, Universit\'{e} Lyon 1, CNRS/IN2P3, IPN-Lyon, Villeurbanne, France
\item \Idef{org131}V.~Fock Institute for Physics, St. Petersburg State University, St. Petersburg, Russia
\item \Idef{org132}Variable Energy Cyclotron Centre, Kolkata, India
\item \Idef{org133}Warsaw University of Technology, Warsaw, Poland
\item \Idef{org134}Wayne State University, Detroit, Michigan, United States
\item \Idef{org135}Wigner Research Centre for Physics, Hungarian Academy of Sciences, Budapest, Hungary
\item \Idef{org136}Yale University, New Haven, Connecticut, United States
\item \Idef{org137}Yonsei University, Seoul, South Korea
\item \Idef{org138}Zentrum f\"{u}r Technologietransfer und Telekommunikation (ZTT), Fachhochschule Worms, Worms, Germany
\end{Authlist}
\endgroup

      %%%%%%% get the latest version before submitting
\else
\ifbibtex
\bibliographystyle{utphys}
\bibliography{biblio}{}
\else
\input{refpreprint.tex}
\fi
\fi
\else
\iffull
\vspace{0.5cm}

\input{refpaper.tex}
\else
\ifbibtex
\bibliographystyle{utphys}
\bibliography{biblio}{}
\else
\input{refpaper.tex}
\fi
\fi
\fi
%==========================================================%
\end{document}